\documentclass[useAMS]{mn2e}
\usepackage{array}
\usepackage{epsfig}
\usepackage{lscape}
\usepackage{footnote}
\usepackage{graphics}
\usepackage{etoolbox}
\usepackage{leftidx}
\usepackage{footnote}
\usepackage{url}

\begin{document}

\title[Line of sight towards QSO PKS~0237$-$233]{An investigation of the line of sight
towards QSO PKS~0237$-$233\thanks{Based on observations carried out
at the European Southern Observatory with the Ultraviolet and
Visible Echelle Spectrograph (UVES) in the course of the large
program 'The Cosmic Evolution of the IGM' 166.A-0106 on the 8.2-m
Very Large Telescope (VLT) Kueyen operated at Paranal Observatory,
Chile.}}
\author[H. Fathivavsari et al.]
  {H.~Fathivavsari$^1$,
  P.~Petitjean$^2$, C.~Ledoux$^3$, P.~Noterdaeme$^2$,
  R.~Srianand$^4$,
  \newauthor 
  H.~Rahmani$^4$,
  A.~Ajabshirizadeh$^{1,5}$\\
  $^1$Department of Theoretical Physics and Astrophysics, University of Tabriz, Tabriz 51664, Iran\\
  $^2$Universit\'e Pierre et Marie Curie - CNRS, UMR\,7095, Institut d'Astrophysique de Paris, 98bis Boulevard Arago, 75014 Paris,  France\\
  $^3$European Southern Observatory, Alonso de Cordova 3107, Vitacura, Casilla 19001, Santiago 19, Chile \\
  $^4$Inter-University Centre for Astronomy and Astrophysics, Post Bag 4, Ganeshkhind, Pune 411 007, India\\
  $^5$Research Institute for Astronomy and Astrophysics of Maragha, Maragha 55134-441, Iran}

\date{ 29 July 2013 }

\pagerange{\pageref{firstpage}--\pageref{lastpage}} \pubyear{2013}

\maketitle

\label{firstpage}

\begin{abstract}
We present a detailed analysis of absorption systems along the line
of sight towards QSO PKS~0237$-$233 using a high resolution spectrum
of signal-to-noise ratio (SNR) $\sim$60-80 obtained with the
Ultraviolet and Visual Echelle Spectrograph mounted on the Very
Large Telescope. This line of sight is known to show a remarkable
overdensity of C~{\sc iv} systems that has been interpreted as
revealing the presence of a supercluster of galaxies. A detailed
analysis of each of these absorption systems is presented. In
particular, for the $z_{\rm abs} =$ 1.6359 (with two components of
log~$N_{\rm H I}$[ cm$^{-2}$]~=18.45, 19.05) and $z_{\rm abs} =$
1.6720 (log~$N_{\rm H I}$~=~19.78) sub-Damped Ly$\alpha$ systems
(sub-DLAs), we measure accurate abundances (resp.
[O/H]~=~$-$1.63$\pm$0.07 and [Zn/H]~=~$-$0.57$\pm$0.05 relative to
solar). While the depletion of refractory elements onto dust grains
in both sub-DLAs is not noteworthy, photoionization models show that
 ionization effects are important in a part of the absorbing gas
of the sub-DLA at $z_{\rm abs} =$ 1.6359 (H~{\sc i} is 95 percent
ionized) and in part of the gas of the sub-DLA at $z_{\rm abs} =$
1.6359 The C~{\sc iv} clustering properties along the line of sight
is studied in order to investigate the nature of the observed
overdensity. We conclude that despite the unusually high number of
C~{\sc iv} systems detected along the line of sight, there is no
compelling evidence for the presence of a single unusual overdensity
 and that the situation is consistent with chance coincidence.

\end{abstract}


 \begin{table*}
\begin{minipage}{22.3cm}
\caption{Solar abundances taken from Asplund et al. (2009).}

\setlength{\tabcolsep}{11.8pt}
\renewcommand{\arraystretch}{1.2}
\begin{tabular}{ >{\scriptsize}c >{\scriptsize}c >{\scriptsize}c >{\scriptsize}c >{\scriptsize}c >{\scriptsize}c >{\scriptsize}c >{\scriptsize}c >{\scriptsize}c >{\scriptsize}c
>{\scriptsize}c >{\scriptsize}c}

\hline \hline

\multicolumn{12}{c}{Solar Abundances in number relative to hydrogen} \\
\hline
  Species  &     &  O  &  Si &  Mg &  S &  Fe &  Cr &  Zn &  Mn &  Ni &    \\
    \hline
  Log Abundance         &     &       $-$3.31         &         $-$4.49   &      $-$4.40        &     $-$4.88        &     $-$4.50         &     $-$6.36         &     -7.44           &     -6.57           &     -5.78           &    \\
\hline

\end{tabular}
\renewcommand{\footnoterule}{}
  \end{minipage}
\end{table*}

 \begin{table*}
\begin{minipage}{22.3cm}
\caption{Elemental column densities in the $z_{\rm abs} = 1.6359$
system.}

\setlength{\tabcolsep}{2.0pt}
\renewcommand{\arraystretch}{1.2}
\begin{tabular}{ >{\scriptsize}c >{\scriptsize}c >{\scriptsize}c >{\scriptsize}c >{\scriptsize}c >{\scriptsize}c >{\scriptsize}c >{\scriptsize}c >{\scriptsize}c >{\scriptsize}c
>{\scriptsize}c >{\scriptsize}c}

\hline \hline

\multicolumn{12}{c}{Low-ion column densities} \\
\hline \hline

  $z$    & $\Delta$V (km~s$^{-1}$) & $b$(km~s$^{-1}$) & log$N$(Mg~{\sc ii}) & log$N$(Mg~{\sc i}) & log$N$(O~{\sc i}) & log$N$(Fe~{\sc ii}) & log$N$(Si~{\sc ii}) & log$N$(Al~{\sc ii}) & log$N$(Al~{\sc iii}) & log$N$(C~{\sc ii}) & log$N$(H~{\sc i})\\

\hline
\multicolumn{12}{c}{Region -- R1} \\
\hline

1.63568 &  $-$22.7 &  10.7$\pm$0.7  & 11.71$\pm$0.01 &     $\leq$10.21  &  $\leq$12.86  &   $\leq$11.22  &   $\leq$12.22  & 10.95$\pm$0.07   &  $\leq$10.90  & ....\footnote{Blended with some features.}\\
1.63579 &  $-$10.2 &   2.3$\pm$0.2  & 11.94$\pm$0.00 & 9.66$\pm$0.21 & 12.77$\pm$0.04 & 11.26$\pm$0.04   & 11.95$\pm$0.13 & 11.25$\pm$0.02   & 10.80$\pm$0.06 & ....$^{a}$\\
1.63588 &    0.0 &   4.2$\pm$0.0  & 12.86$\pm$0.00 & 10.83$\pm$0.02 & 13.23$\pm$0.01 & 12.23$\pm$0.00  & 13.01$\pm$0.01 & 12.01$\pm$0.01   & 11.74$\pm$0.01 & ....$^{a}$        &    18.45$\pm$0.05\\
1.63604 &  +18.2 &   5.0$\pm$0.2  & 11.76$\pm$0.00 &      $\leq$9.93   & 12.73$\pm$0.05 & $\leq$10.43  & 11.67$\pm$0.26 & 11.36$\pm$0.02   & 10.96$\pm$0.05 & ....$^{a}$\\
1.63639 &  +58.0 &   4.3$\pm$1.2  & 10.79$\pm$0.05 &      $\leq$9.68  &   $\leq$11.98   &  $\leq$10.73 &  $\leq$11.79   & 10.52$\pm$013    & 10.49$\pm$0.14 & ....$^{a}$\\
\hline
\multicolumn{12}{c}{Region -- R2} \\
\hline
1.63691 &  +117.1 &   2.3$\pm$0.3  & 11.48$\pm$0.01 &   $\leq$9.82  & 12.33$\pm$0.09 & 11.26$\pm$0.04 & 12.20$\pm$0.07   & 10.51$\pm$0.14   &  $\leq$10.30    & 12.73$\pm$0.06\\
1.63701 &  +128.5 &   1.7$\pm$0.8  & 11.52$\pm$0.01 & 9.81$\pm$0.14 & 12.25$\pm$0.12 & 11.25$\pm$0.04 & 12.05$\pm$0.10   & 10.88$\pm$0.09   &  $\leq$10.15   & 12.83$\pm$0.07\\
1.63712 &  +141.0 &   5.5$\pm$0.4  & 12.36$\pm$0.00 & 10.81$\pm$0.03 & 13.65$\pm$0.01 & 12.37$\pm$0.01 & 12.72$\pm$0.05  & 10.86$\pm$0.34   & 10.63$\pm$0.18  & 13.52$\pm$0.05      \\
1.63717 &  +146.7 &  15.0$\pm$0.7  & 12.85$\pm$0.00 & 10.71$\pm$0.06 & 13.79$\pm$0.01 & 12.65$\pm$0.01 & 13.30$\pm$0.02  & 12.05$\pm$0.01   & 10.91$\pm$0.14  & 14.01$\pm$0.02    &   19.05$\pm$0.05   \\
1.63733 &  +164.9 &   4.8$\pm$0.8  & 11.83$\pm$0.01 & 10.04$\pm$0.11 & 12.67$\pm$0.06 & 11.40$\pm$0.04 & 12.38$\pm$0.06  & 11.24$\pm$0.03   &  $\leq$10.53    & 13.05$\pm$0.04\\
1.63744 &  +177.4 &   6.0$\pm$0.2  & 12.15$\pm$0.00 & 10.46$\pm$0.04 & 12.64$\pm$0.06 & 11.78$\pm$0.01 & 12.42$\pm$0.05  & 11.45$\pm$0.02   & 10.56$\pm$0.13  & 13.25$\pm$0.02\\
1.63759 &  +194.4 &   3.2$\pm$1.5  & 10.76$\pm$0.05 &      $\leq$9.75  & 11.69$\pm$0.43 & 10.45$\pm$0.26 &  $\leq$11.37  & 10.50$\pm$0.13   &   $\leq$10.26   & ....$^{a}$\\
1.63776 &  +213.7 &   3.9$\pm$0.6  & 11.28$\pm$0.02 &  $\leq$9.56  & $\leq$11.92    & 10.86$\pm$0.11 & 12.01$\pm$0.12    & 10.74$\pm$0.08   &   $\leq$10.02      & ....$^{a}$\\
1.63785 &  +224.0 &   6.9$\pm$0.4  & 11.85$\pm$0.01 &   $\leq$9.93 & 12.29$\pm$0.10 & 11.49$\pm$0.03 & 12.47$\pm$0.05    & 11.19$\pm$0.03& $\leq$10.24      & ....$^{a}$\\

\hline \hline

\multicolumn{12}{c}{High-ion column densities} \\
\hline \hline

    &  $z$  & $\Delta$V~(km~s$^{-1}$) &  Ion~(X) & $b$~(km~s$^{-1}$)  &  log~$N$(X)  & $z$  & $\Delta$V~(km~s$^{-1}$) &  Ion~(X) &   $b$~(km~s$^{-1}$)  & log~$N$(X) \\

\hline

   & 1.63577  & $-$12.5 &  C~{\sc iv}    & 22.9$\pm$1.0     &  12.87$\pm$0.02  &  1.63577  & $-$12.5 &  Si~{\sc iv}    & 22.9$\pm$1.0  & 12.00$\pm$0.04\\
   & 1.63588  & 0.0 &  C~{\sc iv}    & 5.9$\pm$0.4     &  12.17$\pm$0.05   &  1.63588  & 0.0 &   Si~{\sc iv}    & 5.9$\pm$0.4  & 12.13$\pm$0.02\\
   & 1.63605  & +19.3 &  C~{\sc iv}    & 4.5$\pm$0.3     &  12.15$\pm$0.03   &  1.63605  & +19.3 &   Si~{\sc iv}    & 4.5$\pm$0.3  & 12.07$\pm$0.01\\
   & 1.63638  & +56.9 &   C~{\sc iv}    & 4.9$\pm$0.5     &  12.22$\pm$0.06   &  1.63638  & +56.9 &   Si~{\sc iv}    & 4.9$\pm$0.5  & 11.63$\pm$0.03\\
   & 1.63651  & +71.6 &  C~{\sc iv}    & 11.8$\pm$4.2     &  12.01$\pm$0.13   &  1.63651  & +71.6 &  Si~{\sc iv}    & 11.8$\pm$4.2  &  $\leq$11.31\\
   & 1.63685  & +110.3 &  C~{\sc iv}    & 16.0$\pm$1.0     &  12.86$\pm$0.02   &  1.63685  & +110.3 &  Si~{\sc iv}    & 16.0$\pm$1.0  & 10.79$\pm$0.21\\
   & 1.63711  & +139,9 &   C~{\sc iv}    & 7.6$\pm$1.3     &  12.91$\pm$0.26   &  1.63711  & +139.9 &   Si~{\sc iv}    & 7.6$\pm$1.3  & 11.78$\pm$0.36\\
   & 1.63720  & +150.1 &  C~{\sc iv}    & 10.6$\pm$6.0     &  12.65$\pm$0.47   &  1.63720  & +150. &  Si~{\sc iv}    & 10.6$\pm$6.0  & 11.74$\pm$0.39\\
   & 1.63742  & +175.1 &  C~{\sc iv}    & 12.3$\pm$1.1     &  12.19$\pm$0.04   &  1.63742  & +175.1 &  Si~{\sc iv}    & 12.3$\pm$1.1  & 12.09$\pm$0.02\\

\hline

\end{tabular}
\renewcommand{\footnoterule}{}
  \end{minipage}
\end{table*}

\section{Introduction}

Metal absorption lines seen in the spectra of high redshift quasars
are thought to be produced by gas clouds associated in some way with
galaxies or their progenitors. This hypothesis is supported in
particular by the amplitude and scale of their clustering which are
consistent with those expected from galaxies (e.g. Scannapieco et
al. 2006). A few lines of sight have been known for long to contain
an unusually large number of absorption systems and the reason for
these puzzling observations have never been fully elucidated.

One instance of such superclustering is seen towards the two quasars
Tol 1037$-$2703 and Tol 1038$-$2712 (Jakobsen et al. 1986) which
have an angular separation of 17.9 arcmin in the plane of the sky,
corresponding to a proper separation of 4.4 $h^{-1}$ Mpc at
$z\sim2$. The spectra of the quasars each exhibit at least five
C~{\sc iv} absorption complexes over the narrow redshift range $1.88
\leq z \leq 2.15$, representing a highly significant overdensity in
the number of absorbers above that expected from Poisson statistics
(Dinshaw $\&$ Impey 1996). One complex lies at the same redshift
along both QSO lines of sight and the rest are coincident to within
$v \leq 2000$ km~s$^{-1}$. The fact that there are similar
absorption features at the same redshift in the spectra of both
these quasars signals that the two lines of sight may actually be
probing the same absorbing structure. The preferred explanation for
the overdensity of C~{\sc iv} absorption systems is that the two
lines of sight are passing through material associated with an
intervening supercluster (Jakobsen et al. 1986; Sargent et al. 1987;
Lespine \& Petitjean 1997; Srianand \& Petitjean 2001) but this
concentration of objects has never been confirmed directly. It is
interesting to note that C~{\sc iv} clustering properties are very
sensitive to the choice of the column density threshold (Scannapieco
et al., 2006).

Another example of an overdensity of absorption systems is observed
along the line of sight to the quasar PKS~0237$-$233. This quasar
was first discovered and studied by Arp, Bolton \& Kinman (1967).
Its absorption spectrum has been the subject of many studies over
the years ( Burbidge 1967; Greenstein et al. 1967; Burbidge et al.
1968; Bahcall et al. 1968; Boksenberg et al. 1975) with three main
complexes at $z_{\rm abs}$~=~1.596, 1.657, 1.674. Foltz et al.
(1993) searched the field for other QSOs to provide background
sources against which the presence of absorption at the same
redshifts could be investigated. They concluded that the complex can
be interpreted as a real spatial overdensity of absorbing clouds
with a transverse size comparable to its extent along the line of
sight, that is of the order of 30~Mpc.

Heisler et al. (1989) found significant clustering signal in the
distribution of C~{\sc iv} systems  out to velocities of $\Delta v
\leq  10,000$ km~s$^{-1}$ in a sample of 55 QSOs observed by Sargent
et al. (1988). They noted that the clustering signal is dominated by
a single large supercluster along the line of sight to
PKS~0237$-$233 spanning a  redshift range from $z = 1.5959$ to $z =
1.6752$. More recently, Scannapieco et al. (2006) studied the line
of sight correlation function of C~{\sc iv} systems using nineteen
lines of sight observed with the Ultraviolet and Visual Echelle
Spectrograph on Very Large Telescope (VLT/UVES). Their sample may
not be large enough to conclude about the clustering signal beyond
500~km~s$^{-1}$. Surprisingly, the redshift evolution of the C~{\sc
iv} systems has not been studied from large samples provided by e.g.
SDSS contrary to what has been done for Mg~{\sc ii} systems. This is
probably related to the difficulties in robustly detecting C~{\sc
iv} systems at intermediate resolution because the two lines of the
doublet are partially blended. Only samples based on high resolution
observations are available (see D'Odorico et al. 2010).

Lines of sight with high overdensities of absorption systems are in
any case rare incidences that are worth investigating in more
details. In this paper, we study the line of sight towards
PKS~0237$-$233 in great detail, using a high spectral resolution ($R
= 45000$) and high signal-to-noise ratio ($S/N \sim$ 60 -- 80 )
spectrum taken with VLT/UVES and taken during the European Southern
Observatory (ESO) Large Program "Cosmic Evolution of the
Intergalactic Medium (IGM)" (Bergeron et al. 2004).

Observations are described in Section~2. Individual systems are
discussed in Section~3 and the Appendix. Results of fits are
analysed in Section~4 using photoionization models. The clustering
properties of C~{\sc iv} absorbers are presented in Section~5 and
conclusions are drawn in Section~6.

\section{Observations}

The spectrum of PKS~0237$-$233 used for this study is of the highest
SNR and spectral resolution. It  was obtained using the Ultraviolet
and Visible Echelle Spectrograph (UVES, Dekker et al. 2000) mounted
on the ESO Kueyen 8.2-m telescope at the Paranal observatory in the
course of the ESO-VLT large programme 'The Cosmic Evolution of the
IGM' (Bergeron et al. 2004). PKS~0237$-$233 was observed through a
1.0 arcsec slit for $\sim$12 hours with dichroic \#1 with central
wavelengths adjusted at 3460 and 5800 \textup{\AA} in the blue and
red arms, respectively, and for another $\sim$14 hours with dichroic
\#2 with central wavelengths at 4370 and 8600~\textup{\AA} in the
blue and red arms respectively. The raw data were reduced using the
UVES pipeline\footnote{
\url{http://www.eso.org/sci/facilities/paranal/instruments/uves/doc/}}.
Individual exposures were air-vacuum corrected and placed in an
heliocentric restframe. Co-addition of the exposures was performed
using a sliding window and weighting the signal by the errors in
each pixel. Great care was taken in computing the error spectrum
while combining the individual exposures. The final combined
spectrum covers the wavelength range 3000 -- 10,000 \textup{\AA}. A
typical SNR$\sim$60 -- 80 per pixel (of 0.035 \textup{\AA}) is
achieved over the whole wavelength range of interest. The detailed
quantitative description of data calibration is presented in Aracil
et al. (2004) and Chand et al. (2004, 2006). We will use these
superb data to make a detailed analysis of the line of sight.
\par\noindent
In the following, we used the solar abundances (photospheric
abundances), log~(X/H)$_{\odot}$, from Asplund et al. (2009) listed
in Table~1 and the metallicity relative to solar of species X,
[X/H]~=~log~X/H~$-$~log~(X/H)$_{\odot}$.

\section{Discussion of individual systems}

This section presents the analysis of absorption profiles for
several systems we chose to study in greater detail. The description
of the other systems detected in the spectrum of PKS~0237$-$233 can
be found in Appendix A. \par\noindent To identify the absorption
systems, we searched first for Mg~{\sc ii} and C~{\sc iv} doublets.
We then identified all metal absorption associated with these
systems. Finally, we checked that there is no system left
unidentified by this procedure. Overall, we identify 18 absorption
systems along this line of sight, three of which are sub-damped
Ly$\alpha$ (sub-DLA) systems at $z_{\rm abs}$$\sim$1.36, 1.63 and
1.67. Sub-DLAs are defined as absorption systems with N(H~{\sc i})
ranging from 10$^{19}$ to 2 $\times$ 10$^{20}$ (P\'eroux et al. 2003
\& Dessauges-Zavadsky et al. 2003).

We use the VPFIT \footnote{
\url{http://www.ast.cam.ac.uk/~rfc/vpfit.html}} package to decompose
the absorption lines into multiple Voigt profile components. The
VPFIT package is a least-square program which minimizes the
${\chi}$$^{2}$ when adjusting a multiple Voigt profile model to
absorption features. The wavelengths and oscillator strengths are
taken from Morton (2003). When fitting the low ion species (O~{\sc
i}, C~{\sc ii}, Si~{\sc ii}, Mg~{\sc ii}, Fe~{\sc ii}), we assumed
that they all have the same kinematic structure which means that
they arise from the same components having the same Doppler
parameters. For the C~{\sc iv} and Si~{\sc iv} profiles, we kept
Doppler parameters independent because we noticed that even though
their absorption profiles correlate very well, C~{\sc iv} can have
broader lines especially in complex profiles (see below and also Fox
et al. 2007a,b).

If an absorption is not detected at the wavelength expected from the
presence of other species in the same system, a 3~${\sigma}$ upper
limit is determined. The redshifts of the H~{\sc i} components are
fixed to that of metals in the case of sub-DLA systems while in the
case of other systems redshifts are considered free parameters.

\begin{figure*}
\centering
\includegraphics[bb=54 358 564 721,clip=,width=0.7\hsize]{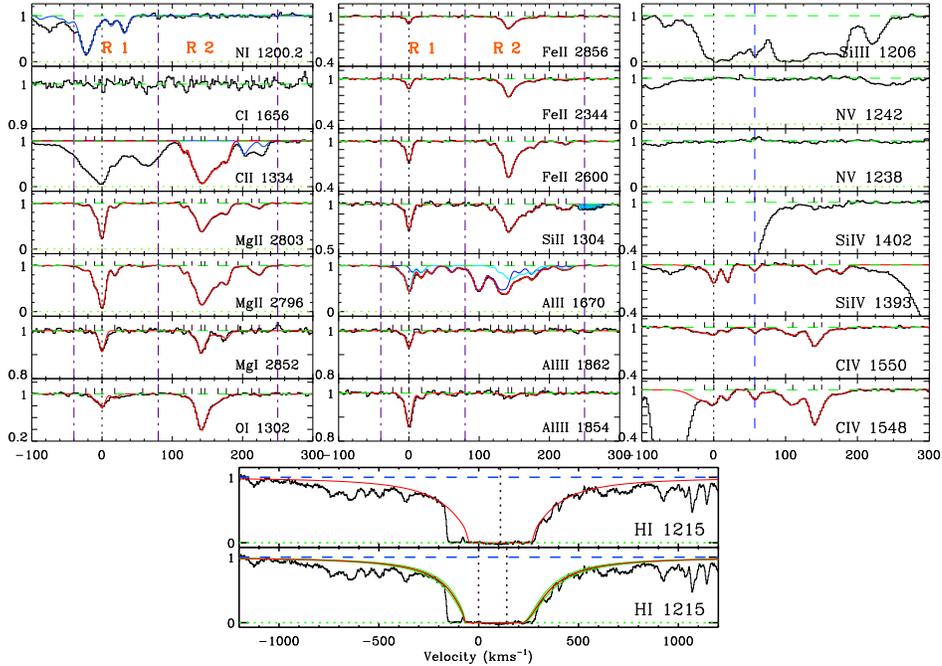}
\caption{Velocity profiles and VPFIT solutions of the species
detected in the sub-DLA at $z_{\rm abs} =$ 1.6359. The blue curves
are VPFIT solutions of absorption from other systems, and the blue
shaded region indicates blends with some intervening absorption.
 Parameters of the fit can be found in Table~2. As illustrated
in the lower panel, the single component fit to the damping wings
was conducted  with the redshift set at $z=1.63658$ while the
2-component fit was performed with the redshifts fixed to those of
the two metal subsystems observed at $z=1.63588$ and $z=1.63717$. As
can be seen in the figure, the latter solution fits the Ly$\alpha$ profile much better.}
\end{figure*}

\subsection{$z_{\rm abs}$ = 1.6359 }

This absorber is a sub-DLA system associated with a number of high
and low ionization species spanning more than 200~km~s$^{-1}$
including C~{\sc ii}, O~{\sc i}, Mg~{\sc ii}, Mg~{\sc i}, Al~{\sc
ii}, Al~{\sc iii}, Fe~{\sc ii}, Si~{\sc ii}, C~{\sc iv} and Si~{\sc
iv}. The velocity profiles and VPFIT solutions (where applicable) of
the H~{\sc i}, low and high ion species are illustrated in Fig.~1.
The measurements are given in Table~2.

In the following the origin of the velocities are set at ${z =
1.63588}$. The metal absorption features in this sub-DLA are seen in
two sub-systems, one between [-40, +80] km~s$^{-1}$ (region R1), and
another between [+80, +250] km~s$^{-1}$ (region R2). The low ion
absorption lines clearly indicate that the bulk of the neutral gas
is located at $v \approx 0$ km~s$^{-1}$ and $v \approx +140$
km~s$^{-1}$ ($z=1.63717$). A damping profile fitted to the
Lyman-$\alpha$ absorption line, with the redshift at $z=1.63685$
(1-component fit), yields a satisfactory fit to the damping wings
for $N_{\rm HI} = 1.58 \times 10^{19}$ cm$^{-2}$ and $b = 46.0$~
km~s$^{-1}$. However, not only the Doppler parameter seems large for
a damped system but also part of the absorption is not accounted for
at $v = -80$~km~s$^{-1}$ (upper panel of the representation of
H~{\sc i}$\lambda$1215 in Fig.~1). We therefore tried to conduct a
2-component fit to the Ly$\alpha$ profile, fixing the redshifts of
the two components to the redshifts of the two metal sub-systems
observed at $z=1.63588$ and $z=1.63717$. As can be seen in Fig.~1
(lower panel of the representation of H~{\sc i}$\lambda$1215), the
2-component solution fits the Ly$\alpha$ profile much better than
the single component one. The parameters of the 2-component solution
are listed in Table~2 (last column). The two components have column
densities of log~$N_{\rm H I}~=~18.45$ and 19.05, respectively.

The metal lines are weak and their fit yields robust column
densities. Using the Mg~{\sc ii} and Fe~{\sc ii} absorption
profiles, we identified 14 velocity components for the low ion
species. The fit to the Al~{\sc ii} absorption profile (cyan curve
in Fig.~1) was performed using the template obtained  on Fe~{\sc ii}
because it is blended with Al~{\sc iii} (blue curve) at $z_{\rm
abs}$~=~1.3647. As can be seen from Fig.~1, the C~{\sc ii}
absorption profile as well as the velocity region where N~{\sc
i}$\lambda1200.2$ is expected, are blended with some absorption
features in region R1. The N~{\sc i} velocity region $-100 \leq v
\leq +80$ km~s$^{-1}$ is contaminated by the Si~{\sc
ii}$\lambda1190$ absorption of a system at $z_{\rm abs} = 1.6574$,
and the VPFIT solution of this Si~{\sc ii} transition is
over-plotted on the observed data as a blue curve. In region R2, the
C~{\sc ii} profile also appears to be blended with the Ni~{\sc
ii}$\lambda1317$ transition of a sub-DLA at $z_{\rm abs} = 1.6720$,
and the blue curve is the VPFIT solution of this Ni~{\sc ii}
transition over-plotted on the C~{\sc ii} profile. We could however
perform a fit of these two profiles in the R2 region.

Absorption by highly ionized gas in this absorber is seen from the
C~{\sc iv} and Si~{\sc iv} doublets. N~{\sc v} is not detected and
O~{\sc vi}  falls outside our wavelength range. The fit to the high
ion species were conducted simultaneously with 9 absorption
components (Fig.~1). The Si~{\sc iv}$\lambda1402$ absorption was
excluded from the fit due to severe blending with the Ly$\alpha$
absorption profile of a system at $z_{\rm abs} = 2.0422$. Moreover,
the C~{\sc iv}$\lambda1548$ profile also appears to be slightly
blended in the blue with the Si~{\sc ii}$\lambda1526$ of a sub-DLA
at $z_{\rm abs} = 1.6720$. However, the fit was reasonably
successful, yielding $\chi_{\nu}^2 =$ 1.25. All C~{\sc iv} and
Si~{\sc iv} measurements are summarized in Table~2.

\begin{figure*}
\centering
\begin{tabular}{c}
\includegraphics[bb=133 367 461 706,clip=,width=0.9\hsize]{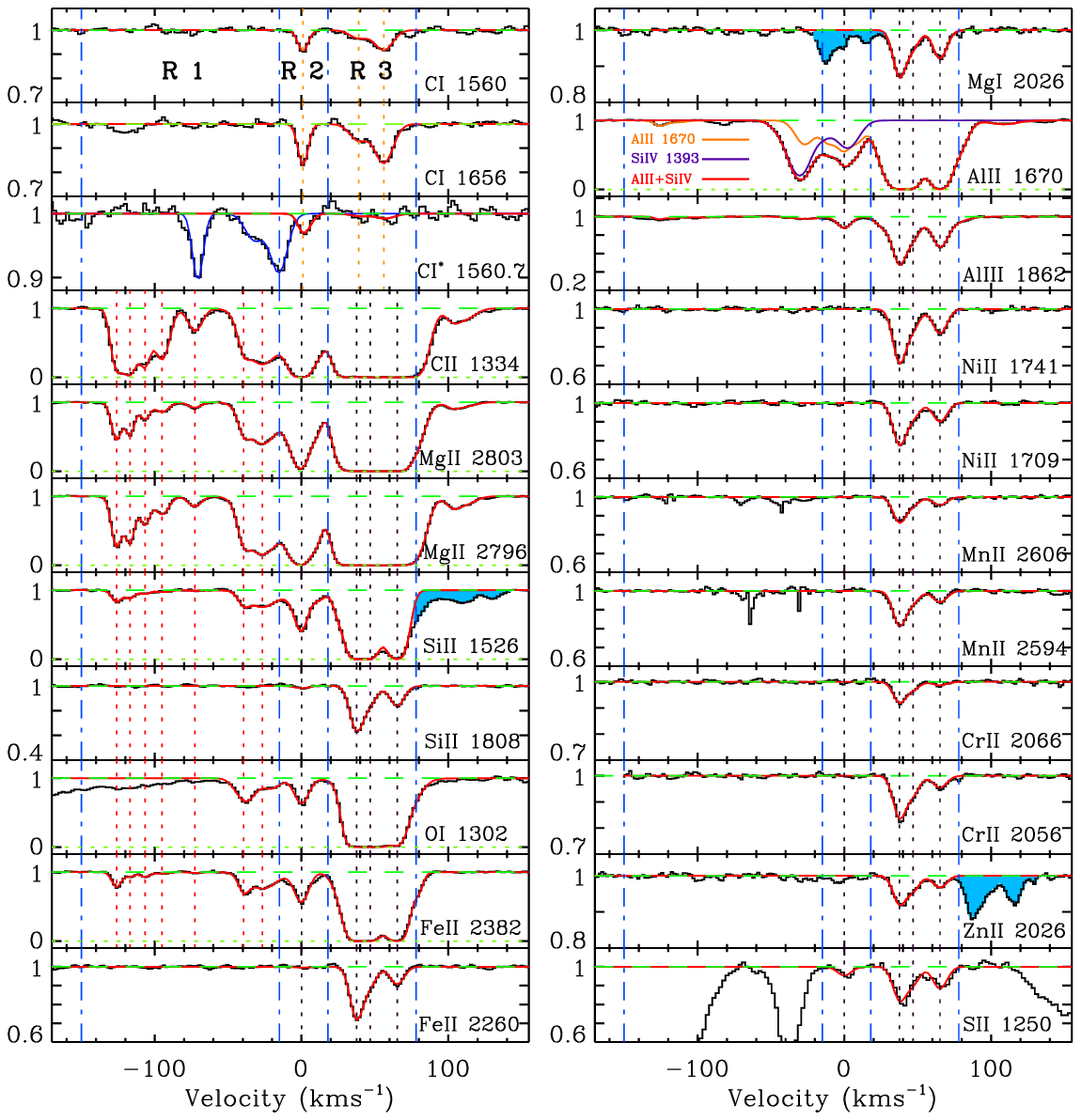}\\
\includegraphics[bb=90 549 540 701,clip=,width=0.9\hsize]{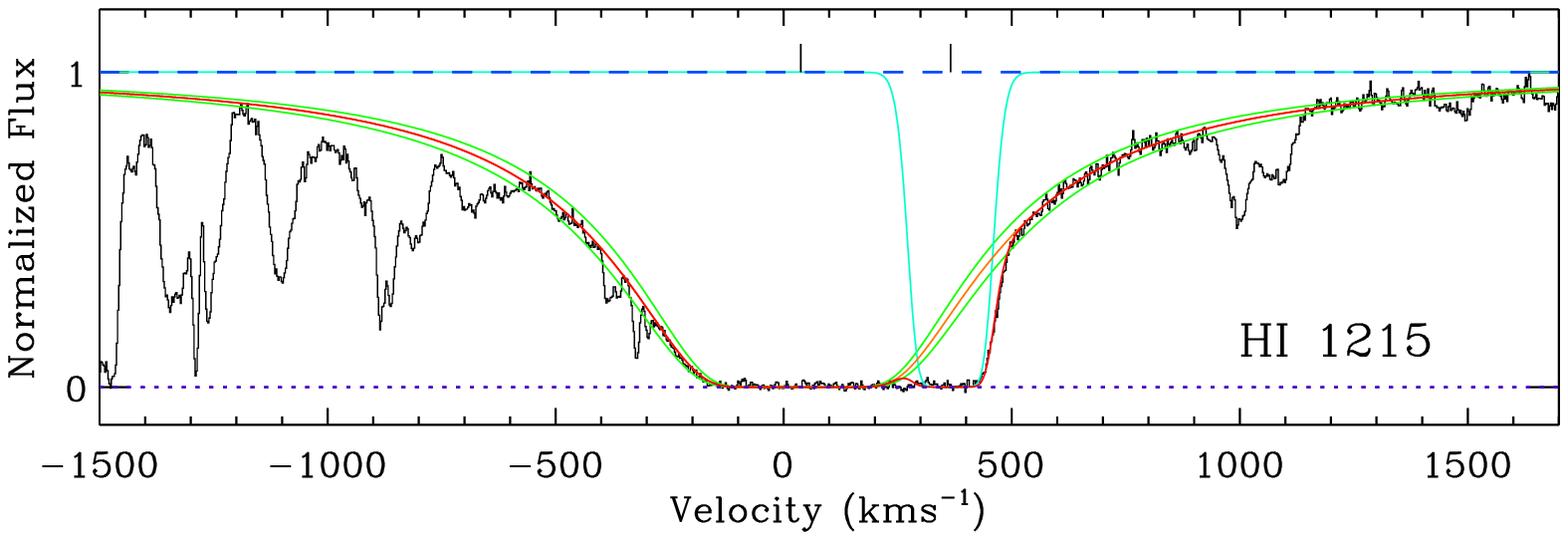}\\
\end{tabular}
\caption{The same as Fig.~1 but for $z_{\rm abs} =$ 1.6720 sub-DLA. Parameters of the fit can be found in Table~3.}
\end{figure*}

\subsection{$z_{\rm abs}$ = 1.6720 }

This absorber is also a sub-DLA in which we detect over 40 metal
lines from 20 different species. A striking feature of this sub-DLA
is the detection of the C~{\sc i} multiplet. The velocity profiles
of H~{\sc i} and some of the neutral and singly ionized species seen
in this system together with a multi-component Voigt profile fit are
shown in Fig.~2. The low-ion species extend over $\simeq 260$
km~s$^{-1}$ in velocity space. Voigt profile fitting to the DLA
absorption profile with redshift fixed to that of the strongest
low-ion component gives log~$N_{\rm H I} = 19.78\pm0.05$. The orange
and green curves overplotted on the data in the lower panel of
Fig.~2 show the Voigt profile fits of this DLA profile. Note that,
in this panel, the red curve also contains the contribution of
another Ly$\alpha$ absorption profile evident at $+300 \leq v \leq
+420$ km~s$^{-1}$ (cyan curve) with log~$N_{\rm H I} = 15.0\pm0.10$.
The redshift of this extra Ly$\alpha$ component corresponds to that
of a strong C~{\sc iv} absorption at $z = 1.67526$.

To facilitate the discussion of this sub-DLA, we divide the velocity
profiles into 3 distinct velocity ranges marked as R1, R2, and R3.
We identified 4 absorption components for the R2+R3 velocity ranges
based on the fits to the Si~{\sc ii} and Fe~{\sc ii} profiles. In
Fig.~2, the blue vertical dot-dashed lines show the boundary of the
three velocity ranges whereas the black vertical dashed lines
indicate the position of the four absorption components identified
in the R2+R3 velocity range. In the R1 velocity range only C~{\sc
ii}, Mg~{\sc ii}, Si~{\sc ii}, O~{\sc i}, Fe~{\sc ii} and Si~{\sc
iii} are detected. We found that a minimum of 7 individual
components were required to optimally fit the absorption features
evident in R1. The red vertical dashed lines in Fig.~2 indicate the
positions of the absorption components. The  red wing of the Si~{\sc
ii}$\lambda1526$ profile is blended with part of the C~{\sc
iv}$\lambda1548$ profile of a system at $z_{\rm abs} = 1.6359$ (blue
shaded area in Fig.~2). Note that the Si~{\sc ii}$\lambda1526$ and
Fe~{\sc ii}$\lambda2382$ transitions were only used to fit the
absorption visible in the R1 velocity range. The blue shaded area in
the Mg~{\sc i}$\lambda2026$ (resp. Zn~{\sc ii}$\lambda2026$)
velocity panel shows absorption from the Zn~{\sc ii}$\lambda2026$
(resp. Mg~{\sc i}$\lambda2026$) transition of this sub-DLA. As
illustrated in Fig.~2, the Al~{\sc ii}$\lambda1670$ is also blended
with the Si~{\sc iv}$\lambda1393$ profile of a system at $z_{\rm
abs}= 2.2028$, so the fit to the Al~{\sc ii} profile was conducted
by including the contribution of this Si~{\sc iv} absorption. The
parameters of the VPFIT solutions for the low and high ion species
are listed in Table~3. The Mg~{\sc ii}, C~{\sc ii}, Al~{\sc ii}, and
O~{\sc i} profiles are clearly saturated in R2 and R3, especially in
R3, so only lower limits to their column densities are given in
Table~3. Note that our column density upper limits are 3~$\sigma$
values.

Figure~3 shows the apparent column density profiles, $N_{a}(v)$ ,
for the Ni~{\sc ii}, Mn~{\sc ii}, Cr~{\sc ii} and Al~{\sc iii}
transitions of this sub-DLA. There is almost no indication of
saturation in these profiles except may be at  $v = +37$~km~s$^{-1}$
in the Ni~{\sc ii} transitions. Here, the Ni~{\sc ii}$\lambda1751$
optical depth is slightly higher than those of the other two
stronger transitions. This is direct evidence of hidden saturation
corresponding to the situation where the Doppler parameter of the
lines is smaller than the spectral resolution. In that case, the
hidden saturation in Ni~{\sc ii} could lead to underestimate the
column density by only $\approx 0.03$ dex. This is of the order of
the error in the column densities and is therefore not important.
Despite the strength of the Al~{\sc iii} absorption profiles there
is no sign of hidden saturation. Moreover, with a few minor
exceptions, the $N_{a}(v)$ curves for the two Mn~{\sc ii}
transitions match quite well, suggesting no hidden saturation as
well. However, the $N_{a}(v)$ profile of the Cr~{\sc
ii}$\lambda2062$ transition does not coincide with that of the other
two transitions. This is probably due to the blending with some
unidentified absorption features hidden in the profile of Cr~{\sc
ii}$\lambda2062$. Therefore, we will adopt the Cr~{\sc ii} column
density determined from the $\lambda2056$  and $\lambda2066$
features only.

The redshift alignment between the C~{\sc i} multiplet (i.e. C~{\sc
i} and C~{\sc i}$^{*}$) and low ion species is not very good, so the
fit to the C~{\sc i} multiplet was performed separately (see
Fig.~2). The results of the fit are listed in Table~4. Since both
C~{\sc i} and C~{\sc i}$^{*}$ are clearly detected for the component
at $z$ = 1.67201, we calculate the excitation temperature, between
the J = 0 and J = 1 fine structure levels of C~{\sc i} to be $T_{\rm
ex}$ = 11.42 K. This is consistent with but larger than the
predicted value of the Cosmic Microwave Background (CMB) temperature
$T_{\rm CMB}$ = 7.28~K from the standard cosmology. Indeed, C~{\sc
i} fine structure levels can be populated by other excitation
processes such as collision and UV pumping (see e.g. Ge et al. 1997;
Srianand et al. 2000). Moreover, Kanekar et al. (2009) report a
tentative detection of H~{\sc i} 21 cm absorption in this system.
They determined an H~{\sc i} 21 cm integrated optical depth of 0.076
$\pm$ 0.016 and a covering factor of 0.9 which Ellison et al. (2012)
later used to derive a spin temperature of $T_{s}$ = 380 $\pm$ 127
K.

Finally, absorption by highly ionized gas in this sub-DLA is seen in
C~{\sc iv}, Si~{\sc iv}, N~{\sc v} and Si~{\sc iii}. Figure~4 gives
the velocity profiles and VPFIT solutions of these species. The
parameters are listed in Table~3. We chose not to fit the N~{\sc v}
doublet and Si~{\sc iii} profiles due to severe blending and
saturation. Moreover, both transitions of the Si~{\sc iv} doublet
are partly blended with some forest absorption. In the C~{\sc iv}
$\lambda1548$ (resp. C~{\sc iv} $\lambda1550$) velocity panel of
Fig.~4, the blue shaded areas indicate blends with the C~{\sc iv}
$\lambda1550$ (resp. C~{\sc iv} $\lambda1548$) absorption of the
same system.

\begin{figure}
\includegraphics[bb=70 360 558 720,clip=,width=1.01\hsize]{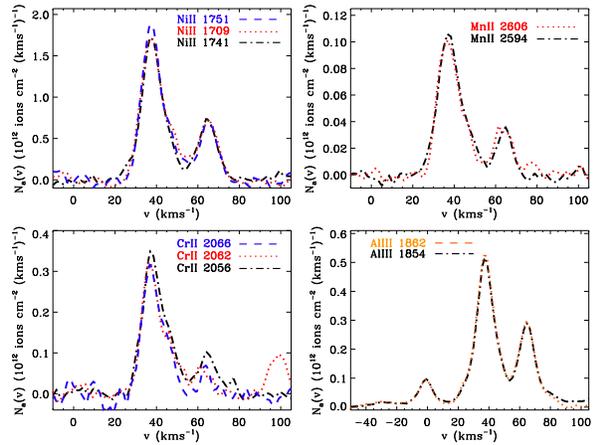}
\caption{Apparent column density analyses of the Ni~{\sc ii},
Mn~{\sc ii}, Cr~{\sc ii} and Al~{\sc iii} absorption profiles in the $z_{\rm abs} =$ 1.6720 sub-DLA.}
\end{figure}

\begin{figure}
\includegraphics[bb=130 360 558 720,clip=,width=1.2\hsize]{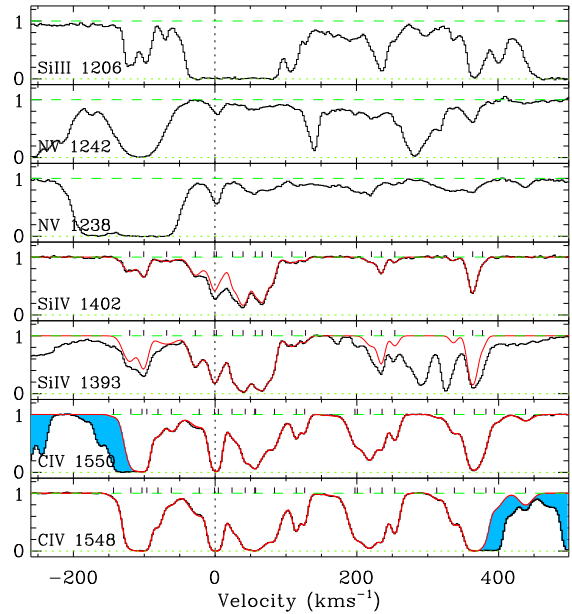}
\caption{Observed velocity profiles and VPFIT solutions of the
high-ion transitions in the $z_{\rm abs} =$ 1.6720 sub-DLA. Blue
shaded regions show blends with other lines, as described in Sect 3.2.
Parameters of the fit can be found in Table~3.}
\end{figure}

 \begin{table*}
\begin{minipage}{22.3cm}

\caption{Elemental column densities for the $z_{\rm abs} = 1.6720$ sub-DLA.
}
\setlength{\tabcolsep}{2.0pt}
\renewcommand{\arraystretch}{1.2}
\begin{tabular}{ >{\scriptsize}c >{\scriptsize}c >{\scriptsize}c >{\scriptsize}c >{\scriptsize}c >{\scriptsize}c >{\scriptsize}c >{\scriptsize}c >{\scriptsize}c >{\scriptsize}c >{\scriptsize}c >{\scriptsize}c}

\hline \hline


\multicolumn{12}{c}{Low-ion column densities} \\
\hline

   $z$          &   1.67088  &  1.67096   &  1.67105    & 1.67115    &  1.67136   &  1.67165   &  1.67176    &  1.67200   &  1.67234   &  1.67242   &  1.67258    \\
\hline
$\Delta$V(km~s$^{-1}$) & $-$125.7 & $-$116.7 &  $-$106.6 & $-$95.4 & $-$71.8 & $-$39.3 & $-$26.9 & 0.0 & +38.1 & +47.1 & +65.1  \\
\hline
$b$(km~s$^{-1}$) & 2.6$\pm$0.1 &  2.6$\pm$0.2 &  2.6$\pm$0.4 & 5.2$\pm$0.5 & 5.1$\pm$0.5 & 3.1$\pm$0.3 & 12.4$\pm$0.3 & 5.4$\pm$0.0 & 4.8$\pm$0.1 & 5.1$\pm$1.1 & 5.5$\pm$0.2  \\

$N$(Mg~{\sc ii}) & 12.51$\pm$0.01 & 12.39$\pm$0.01 & 12.01$\pm$0.01 & 11.88$\pm$0.02 & 11.60$\pm$0.02 & 12.16$\pm$0.01 & 12.97$\pm$0.00 & $\geq$13.35 &   $\geq$13.55 & $\geq$13.55 & $\geq$13.45 \\
$N$(Mg~{\sc i})  & ....           & .... & .... & .... & .... & .... & .... & .... \footnote{Blended with other lines.} &   12.38$\pm$0.01 & 11.94$\pm$0.04 & 12.20$\pm$0.02 \\
$N$(C~{\sc ii})  & 13.95$\pm$0.03 & 14.56$\pm$0.05 & 13.75$\pm$0.03 & 13.55$\pm$0.01 & 13.01$\pm$0.02 & 13.08$\pm$0.03 & 13.93$\pm$0.01 & $\geq$14.55 &   $\geq$14.55 & $\geq$14.55 & $\geq$14.45 \\
$N$(O~{\sc i})   & ....\footnote{Blended with some unidentified forest absorption.} & ....$^{c}$ & ....$^{c}$ & ....$^{c}$ & ....$^{c}$ & 13.34$\pm$0.03 & 13.47$\pm$0.03 & 13.50$\pm$0.02 &   $\geq$14.85 & $\geq$14.85 & $\geq$14.80 \\
$N$(N~{\sc i})   & ....           & .... & .... & .... & .... & .... & .... & .... &   13.35$\pm$0.03 & 12.92$\pm$0.03 & 12.80$\pm$0.03 \\
$N$(Fe~{\sc ii}) & 12.00$\pm$0.02 & 11.43$\pm$0.06 & 11.46$\pm$0.06 & 11.12$\pm$0.15 & 10.82$\pm$0.29 & 12.06$\pm$0.02 & 12.49$\pm$0.01 & 12.52$\pm$0.05 &   14.38$\pm$0.00 & 13.87$\pm$0.02 & 13.89$\pm$0.01 \\
$N$(Si~{\sc ii}) & 12.46$\pm$0.05 & 12.28$\pm$0.07 & 11.92$\pm$0.14 & 11.78$\pm$0.22 & 11.21$\pm$0.83 & 12.46$\pm$0.05 & 13.08$\pm$0.02 & 13.33$\pm$0.04 &   14.69$\pm$0.00 & 14.23$\pm$0.01 & 14.33$\pm$0.01 \\
$N$(Ni~{\sc ii}) & ....           & .... & .... & .... & .... & .... & .... & $\leq$11.73 &   13.26$\pm$0.01 & 12.77$\pm$0.02 & 12.92$\pm$0.01 \\
$N$(Mn~{\sc ii}) & ....           & .... & .... & .... & .... & .... & .... & $\leq$10.78 &   12.05$\pm$0.01 & 11.54$\pm$0.02 & 11.58$\pm$0.02 \\
$N$(Cr~{\sc ii}) & ....           & .... & .... & .... & .... & .... & .... & $\leq$11.11 &   12.52$\pm$0.01 & 12.15$\pm$0.04 & 11.98$\pm$0.07 \\
$N$(Zn~{\sc ii}) & ....           & .... & .... & .... & .... & .... & .... & $\leq$10.50 &   11.51$\pm$0.02 & 11.22$\pm$0.05 & 11.16$\pm$0.05 \\
$N$(S~{\sc ii})  & ....           & .... & .... & .... & .... & .... & .... & 13.49$\pm$0.08 &   14.07$\pm$0.02 & 13.76$\pm$0.05 & 13.91$\pm$0.03 \\
$N$(Al~{\sc ii}) & ....           & .... & .... & .... & .... & .... & .... & 11.91$\pm$0.03 &   $\geq$13.20 & $\geq$13.00 & $\geq$13.10 \\
$N$(Al~{\sc iii})& ....           & .... & .... & .... & .... & .... & .... & 12.05$\pm$0.01 &   12.75$\pm$0.00 & 12.31$\pm$0.01 & 12.41$\pm$0.01 \\


\hline \hline

\multicolumn{12}{c}{High-ion column densities} \\
\hline

    &  $z$  & $\Delta$V~(km~s$^{-1}$) &  Ion~(X) & $b$~(km~s$^{-1}$)  &  log~$N$(X)  & $z$  & $\Delta$V~(km~s$^{-1}$) &  Ion~(X) &   $b$~(km~s$^{-1}$)  & log~$N$(X) \\

\hline
    &  1.67072  & $-$143.6 &  C~{\sc iv}     &  7.0$\pm$2.3   &  12.11$\pm$0.14  &  1.67526  & +365.5 &  C~{\sc iv}     & 11.2$\pm$0.2   &  14.25$\pm$0.01 \\
    &  1.67094  & $-$119.0 &  C~{\sc iv}     & 10.1$\pm$0.6   &  13.91$\pm$0.07  &  1.67541  & +382.3 &  C~{\sc iv}     &  5.2$\pm$1.1   &  13.07$\pm$0.13 \\
    &  1.67108  & $-$103.2 &  C~{\sc iv}     &  9.1$\pm$0.7   &  14.27$\pm$0.05  &  1.67552  & +394.7 &  C~{\sc iv}     & 13.8$\pm$2.5   &  13.02$\pm$0.11 \\
    &  1.67114  & $-$96.5 &  C~{\sc iv}     &  3.6$\pm$1.7   &  13.36$\pm$0.27   &  1.67591  & +438.4 &  C~{\sc iv}     & 11.8$\pm$0.5   &  12.81$\pm$0.01 \\
    &  1.67128  & $-$80.8 &  C~{\sc iv}     &  8.4$\pm$1.3   &  13.25$\pm$0.07   &  1.67093  & $-$120.1 &  Si~{\sc iv}     &  11.6$\pm$0.5  &  12.82$\pm$0.02 \\
    &  1.67145  & $-$61.7 &  C~{\sc iv}     &  9.5$\pm$1.7   &  12.71$\pm$0.08   &  1.67111  & $-$99.9 &  Si~{\sc iv}     &   7.1$\pm$0.4  &  12.78$\pm$0.02 \\
    &  1.67180  & $-$22.4 &  C~{\sc iv}     & 17.2$\pm$1.0   &  13.29$\pm$0.03   &  1.67139  & $-$68.4 &  Si~{\sc iv}     &  19.6$\pm$1.9  &  12.46$\pm$0.03 \\
    &  1.67199  & $-$1.1 &  C~{\sc iv}     &  7.3$\pm$0.3   &  14.06$\pm$0.07    &  1.67175  & $-$28.1 &  Si~{\sc iv}     &  10.3$\pm$0.8  &  12.87$\pm$0.07 \\
    &  1.67204  & +4.5 &  C~{\sc iv}     &  5.3$\pm$0.6   &  13.79$\pm$0.13      &  1.67199  & $-$1.1 &  Si~{\sc iv}     &   4.6$\pm$2.8  &  12.54$\pm$0.47 \\
    &  1.67221  & +23.5 &  C~{\sc iv}     &  7.4$\pm$5.4   &  12.65$\pm$0.64     &  1.67202  & +2.2 &  Si~{\sc iv}     &  15.2$\pm$5.6  &  13.23$\pm$0.02 \\
    &  1.67238  & +42.6 &  C~{\sc iv}     &  4.3$\pm$1.0   &  13.25$\pm$0.12     &  1.67222  & +24.7 &  Si~{\sc iv}     &   4.3$\pm$3.4  &  12.67$\pm$0.50 \\
    &  1.67250  & +56.1 &  C~{\sc iv}     &  6.8$\pm$0.4   &  13.73$\pm$0.02     &  1.67235  & +39.3 &  Si~{\sc iv}     &  10.1$\pm$1.9  &  13.56$\pm$0.06 \\
    &  1.67251  & +57.2 &  C~{\sc iv}     & 30.5$\pm$1.2   &  14.26$\pm$0.03     &  1.67251  & +57.2 &  Si~{\sc iv}     &   4.7$\pm$1.9  &  12.93$\pm$0.34 \\
    &  1.67275  & +84.1 &  C~{\sc iv}     &  5.5$\pm$0.6   &  12.91$\pm$0.04     &  1.67259  & +66.2 &  Si~{\sc iv}     &   6.4$\pm$1.7  &  13.31$\pm$0.12 \\
    &  1.67302  & +114.4 &  C~{\sc iv}     &  5.8$\pm$0.2   &  13.19$\pm$0.02    &  1.67271  & +79.6 &  Si~{\sc iv}     &   6.9$\pm$1.0  &  12.96$\pm$0.09 \\
    &  1.67313  & +126.7 &  C~{\sc iv}     &  4.7$\pm$0.3   &  13.00$\pm$0.02    &  1.67297  & +108.8 &  Si~{\sc iv}     &  12.9$\pm$2.2  &  12.41$\pm$0.05 \\
    &  1.67376  & +197.4 &  C~{\sc iv}     & 39.4$\pm$8.0   &  12.89$\pm$0.24    &  1.67314  & +127.9 &  Si~{\sc iv}     &   4.9$\pm$2.2  &  11.75$\pm$0.17 \\
    &  1.67379  & +200.8 &  C~{\sc iv}     & 12.3$\pm$0.5   &  13.65$\pm$0.04    &  1.67397  & +220.9 &  Si~{\sc iv}     &   9.9$\pm$2.2  &  12.29$\pm$0.11 \\
    &  1.67395  & +218.7 &  C~{\sc iv}     &  9.6$\pm$0.5   &  13.82$\pm$0.02    &  1.67410  & +235.5 &  Si~{\sc iv}     &   5.4$\pm$0.6  &  12.60$\pm$0.05 \\
    &  1.67410  & +235.5 &  C~{\sc iv}     &  6.5$\pm$0.3   &  13.45$\pm$0.03    &  1.67427  & +254.6 &  Si~{\sc iv}     &   6.3$\pm$1.1  &  12.12$\pm$0.04 \\
    &  1.67426  & +253.4 &  C~{\sc iv}     &  7.3$\pm$0.2   &  13.35$\pm$0.02    &  1.67500  & +336.4 &  Si~{\sc iv}     &   5.7$\pm$1.3  &  11.92$\pm$0.06 \\
    &  1.67479  & +312.9 &  C~{\sc iv}     &  7.4$\pm$0.9   &  12.28$\pm$0.05    &  1.67525  & +364.4 &  Si~{\sc iv}     &   7.0$\pm$0.2  &  13.16$\pm$0.01 \\
    &  1.67501  & +337.5 &  C~{\sc iv}     & 11.5$\pm$0.4   &  13.33$\pm$0.01    &  1.67538  & +379.0 &  Si~{\sc iv}     &   5.8$\pm$1.7  &  12.12$\pm$0.11 \\

\hline

\end{tabular}
\renewcommand{\footnoterule}{}
  \end{minipage}
\end{table*}

\subsection{Systems with $z>2$}

 We will single out some of the systems with highest redshifts to determine
whether they are intervening systems or systems associated to the
quasar (see e.g. Petitjean et al. 1992).

\subsubsection{$z_{\rm abs}$ = 2.0422 }

Figure~5 shows the velocity profiles and VPFIT solutions (wherever
applicable) of the Ly$\alpha$, Ly$\beta$ and high ion transitions
(C~{\sc iv}, N~{\sc v} and O~{\sc vi} doublets) regions associated
with this system. The results of the fits are listed in Table~5.
 As illustrated in Fig.~5, it is apparent that the H~{\sc i}
column density of the main complex around 0 km~s$^{-1}$ is very
small. We derive an upper limit of log~$N_{\rm H I}~<~12.90$ (blue
curves in the H~{\sc i} velocity panels). It is interesting to note
that the Si~{\sc iv} doublet is not detected in this system. As
shown in Fig.~5, the O~{\sc vi}$\lambda1037$ absorption is
completely lost in the strong Ly$\alpha$ absorption of the system at
$z_{\rm abs} = 1.5965$. Moreover, the O~{\sc vi}$\lambda1031$
profile is also severely blended with Si~{\sc ii}$\lambda1190$ of
the system at $z_{\rm abs} = 1.6359$. In Fig.~5, the blue curve in
the O~{\sc vi}$\lambda1031$ velocity panel is the VPFIT solution of
this intervening Si~{\sc ii} absorption over-plotted on the observed
data.

A 6-component fit was performed on the C~{\sc iv} doublet profiles.
The 3 absorption components of the C~{\sc iv} doublet in the
velocity range $-120 \leq v \leq -70$ km~s$^{-1}$ are not seen in
the N~{\sc v} doublet. The N~{\sc v}$\lambda1242$ profile around
zero velocity is also partially blended with some unidentified
absorption features, and we chose not to include it in the fit. We
found that a successful solution is achieved for N~{\sc v} by fixing
the Doppler parameters and redshifts of the first and third
components to the values of their corresponding components in the
C~{\sc iv} complex.

 \begin{table}
 \centering
 \caption{Column densities of the C~{\sc i} fine structure levels in the sub-DLA at $z_{\rm abs} = 1.6720$.}
 \setlength{\tabcolsep}{3.6pt}
\renewcommand{\arraystretch}{1.2}
\begin{tabular}{c c c c c c }
\hline\hline

  $z$ & $\Delta$V~(km~s$^{-1}$)   & $b$~(km~s$^{-1}$)  &  log~$N$(C~{\sc i})    &  log~$N$(C~{\sc i}$^{*}$) \\
\hline

1.67201 & +1.1 &  3.4$\pm$0.3 & 12.43$\pm$0.01     & 12.01$\pm$0.02    \\
1.67235 & +39.3 &  7.3$\pm$1.1 & 12.20$\pm$0.06     & $\leq$11.84    \\
1.67250 & +56.1 &  7.8$\pm$0.6 & 12.62$\pm$0.02     & $\leq$11.97    \\

\hline
\end{tabular}
\end{table}

\begin{figure*}
\centering
\includegraphics[bb=44 374 559 708,clip=,width=0.9\hsize]{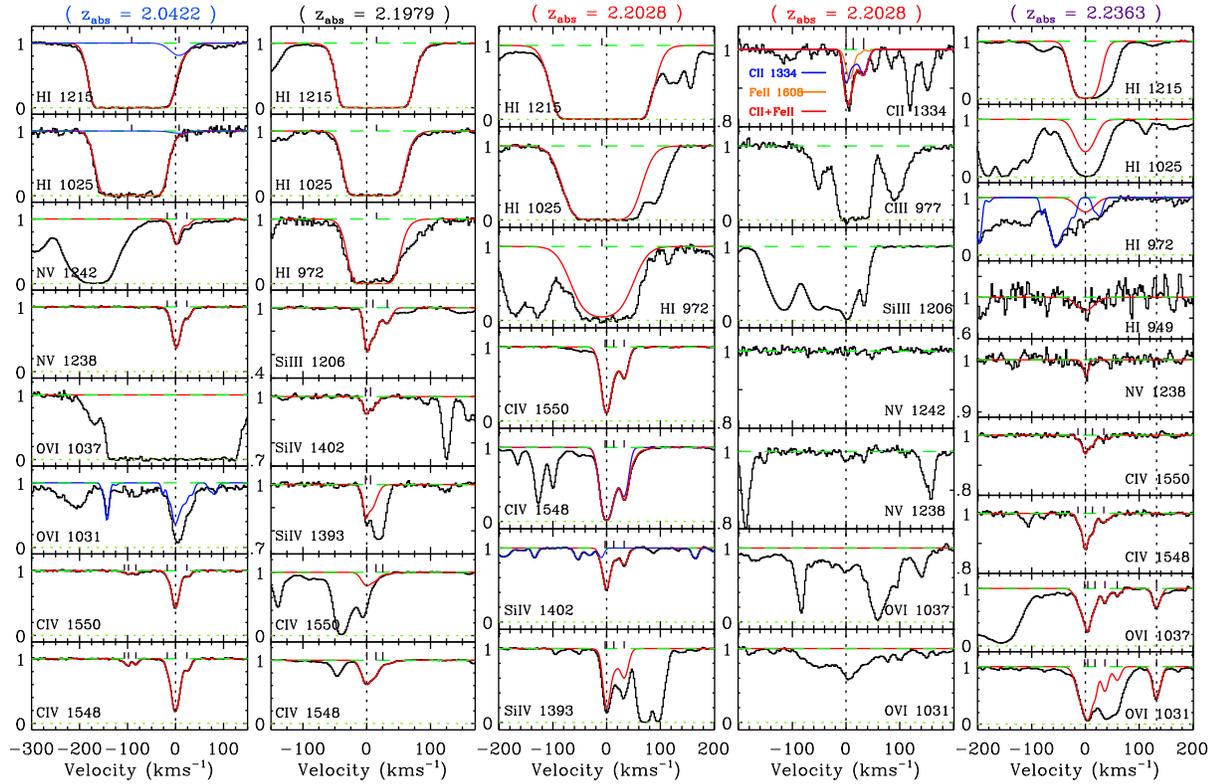}\\
\caption{Observed velocity profiles and VPFIT solutions of the
species
detected in the $z_{\rm abs} =$ 2.0422, 2.1979, 2.2028, and 2.2363 absorbers.
 Parameters of the fit can be found in Table~5.}
\end{figure*}

 \begin{table}
 \centering
\caption{High-ion and H~{\sc i} column densities for the $z_{\rm
abs} =$ 2.0422, 2.1979, 2.2028, and 2.2363 absorbers.}
 \setlength{\tabcolsep}{4.7pt}
\renewcommand{\arraystretch}{1.1}

\begin{tabular}{c c c c c}

\hline\hline

    $z$ & $\Delta$V~(km~s$^{-1}$)  &  Ion~(X) & $b$~(km~s$^{-1}$) & log~$N$(X) \\

\hline
\multicolumn{5}{c}{$z_{\rm abs} = 2.0422$} \\
\hline
2.04224  & +7.1 &  H~{\sc i}     & 25.0   & $\le$12.90\\
2.04108  & $-$107.4 &  C~{\sc iv}     & 8.9$\pm$5.8   & 12.08$\pm$0.55\\
2.04117  & $-$98.6 &  C~{\sc iv}     &   4.6$\pm$3.4   &   12.08$\pm$0.55  \\
2.04133  & $-$82.8 &  C~{\sc iv}     &   7.0$\pm$1.1   &   12.26$\pm$0.04  \\
2.04199  & $-$17.7 &  C~{\sc iv}     &   5.4$\pm$0.6   &   12.38$\pm$0.05  \\
2.04217  & 0.0 &  C~{\sc iv}     &   9.3$\pm$0.1   &   13.58$\pm$0.00  \\
2.04241  & +23.6 &  C~{\sc iv}     &  10.6$\pm$0.7   &   12.72$\pm$0.02  \\
2.04199  & $-$17.7 &  N~{\sc v}     &   5.4$\pm$0.6   &   12.22$\pm$0.08  \\
2.04218  & +1.0 &  N~{\sc v}     &   9.3$\pm$0.1   &   13.55$\pm$0.01  \\
2.04241  & +23.6 &  N~{\sc v}     &  10.6$\pm$0.7   &   12.91$\pm$0.02  \\
\hline
\multicolumn{5}{c}{$z_{\rm abs} = 2.1979$} \\
\hline
2.19802  & +13.1 &  H~{\sc i}   & 46.5$-$32.8   & 15.80$-$16.22\\
2.19788  & 0.0 &  C~{\sc iv}   & 10.2$\pm$0.2   & 13.06$\pm$0.01\\
2.19803  & +14.1 &  C~{\sc iv}   &  6.7$\pm$0.7    &  12.50$\pm$0.09  \\
2.19818  & +28.1 &  C~{\sc iv}   & 15.7$\pm$2.5    &  12.40$\pm$0.09  \\
2.19786  & $-$1.9 &  Si~{\sc iv}   &  2.1$\pm$1.6    &  11.50$\pm$0.13  \\
2.19794  & +5.6 &  Si~{\sc iv}   & 11.9$\pm$0.8    &  12.14$\pm$0.04  \\
2.19787  & $-$1.0 &  Si~{\sc iii}   &  4.6$\pm$0.3    &  11.85$\pm$0.06  \\
2.19798  & +9.4 &  Si~{\sc iii}   & 10.0$\pm$1.0    &  12.02$\pm$0.05  \\
2.19822  & +31.9 &  Si~{\sc iii}   &  8.5$\pm$0.5    &  11.65$\pm$0.02  \\
\hline
\multicolumn{5}{c}{$z_{\rm abs} = 2.2028$} \\
\hline
2.20268 & $-$9.3 &  H~{\sc i} &  49.0$-$45.8      & 15.47$-$15.60 \\
2.20266 & $-$10.3 &  C~{\sc iv} &  8.9$\pm$2.3      & 13.12$\pm$0.38 \\
2.20278 & 0.0 &  C~{\sc iv} &  7.7$\pm$0.7      & 13.93$\pm$0.07 \\
2.20293 & +15.0 &  C~{\sc iv} &  7.7$\pm$1.4      & 13.22$\pm$0.10 \\
2.20312 & +32.8 &  C~{\sc iv} &  9.2$\pm$0.3      & 13.39$\pm$0.02 \\
2.20266 & $-$10.3 &  C~{\sc ii} &  8.9$\pm$2.3      &     11.45$\pm$0.20 \\
2.20278 & 0.0 &  C~{\sc ii} &  7.7$\pm$0.7      &     12.45$\pm$0.20 \\
2.20293 & +15.0 &  C~{\sc ii} &  7.7$\pm$1.4      &     12.09$\pm$0.18 \\
2.20312 & +32.8 &  C~{\sc ii} &  9.2$\pm$0.3      &     12.41$\pm$0.09 \\
2.20266 & $-$10.3 &  Si~{\sc iv} &  8.9$\pm$2.3      &  12.33$\pm$0.36      \\
2.20278 & 0.0 &  Si~{\sc iv} &  7.7$\pm$0.7      &  13.09$\pm$0.07     \\
2.20293 & +15.0 &  Si~{\sc iv} &  7.7$\pm$1.4      &  12.33$\pm$0.12      \\
2.20312 & +32.8 &  Si~{\sc iv} &  9.2$\pm$0.3      &  12.68$\pm$0.02      \\
2.20280 & +2.8 &  N~{\sc v} &  11.4$\pm$1.6      &  12.00$\pm$0.05      \\
2.20312 & +32.8 &  N~{\sc v} &  3.2$\pm$1.3      &  11.70$\pm$0.06      \\
\hline
\multicolumn{5}{c}{$z_{\rm abs} = 2.2363$} \\
\hline
2.23633  & 0.0 &  H~{\sc i}   & 21.9   & 14.18$\pm$0.02\\
2.23618  & $-$13.9 &  C~{\sc iv}   & 5.0$\pm$1.5   & 11.58$\pm$0.16\\
2.23633  & 0.0 &  C~{\sc iv}   &  7.4$\pm$1.2   &  12.38$\pm$0.05 \\
2.23647  & +13.0 &  C~{\sc iv}   &  4.8$\pm$1.2   &  11.91$\pm$0.11 \\
2.23670  & +34.3 &  C~{\sc iv}   & 13.6$\pm$1.4   &  11.98$\pm$0.03 \\
2.23631  & $-$1.9 &  O~{\sc vi}   & 15.5$\pm$5.5   &  13.98$\pm$0.42 \\
2.23637  & +3.7 &  O~{\sc vi}   &  5.8$\pm$2.3   &  13.78$\pm$0.41 \\
2.23652  & +17.6 &  O~{\sc vi}   &  5.5$\pm$4.1   &  13.25$\pm$0.80 \\
2.23672  & +36.1 &  O~{\sc vi}   &  7.7$\pm$1.7   &  13.37$\pm$0.08 \\
2.23697  & +59.3 &  O~{\sc vi}   &  7.8$\pm$1.9   &  13.07$\pm$0.07 \\
2.23776  & +132.4 &  O~{\sc vi}   &  9.0$\pm$0.3   &  13.58$\pm$0.01 \\
2.23634  & +1.0 &  N~{\sc v}    & 6.1$\pm$1.5   & 11.88$\pm$0.06\\

\hline

\end{tabular}
\end{table}

\subsection{$z_{\rm abs}$ = 2.1979 }
We detect the C~{\sc iv} and Si~{\sc iv} doublets together with
Si~{\sc iii}$\lambda1206$, Ly$\alpha$, Ly$\beta$ and Ly$\gamma$
absorption profiles in this system. Figure~5 shows the velocity
profiles and VPFIT solutions for these transitions. Note that the
Ly$\gamma$ profile appears to be blended with some unidentified
forest lines but can still be used to constrain the H~{\sc i} column
density. From these three H~{\sc i} lines we derive
15.80~$<$~log~$N_{\rm H I}$~$<$~16.22. Due to blending, the C~{\sc
iv}$\lambda1550$ and Si~{\sc iv}$\lambda1393$ profiles were excluded
from the Voigt profile fitting. Moreover, because of the lack of
alignment between individual profiles, we chose not to tie the high
ion species during the fitting process. The results of the fit are
listed in Table~5.

\subsection{$z_{\rm abs}$ = 2.2028 }

The velocity profiles and VPFIT solutions for this absorption system
are shown in Fig.~5 and the results are presented in Table~5. We
used Ly$\alpha$ and Ly$\beta$ absorption to derive a lower limit to
the H~{\sc i} column density of log~$N_{\rm H I} \ge$ 15.47. The
Ly$\gamma$ absorption profile although contaminated brings
additional information to derive an upper limit of log~$N_{\rm H I}
\le$ 15.60. There are two C~{\sc ii} transitions in the observed
wavelength range, one of which (C~{\sc ii}$\lambda1036$) is lost in
the forest. We fit the C~{\sc ii}$\lambda1334$ absorption feature
including the contribution of the Fe~{\sc ii}$\lambda1608$
absorption of the system at $z_{\rm abs} = 1.6574$ with which it is
blended (see Fig.~5). The C~{\sc iv} and Si~{\sc iv} absorption
profiles were fitted simultaneously without including the C~{\sc
iv}$\lambda1548$ and Si~{\sc iv}$\lambda1393$ transitions. Indeed,
Si~{\sc iv}$\lambda1393$ is blended with Al~{\sc ii}$\lambda1670$ at
$z_{\rm abs} = 1.6720$ and C~{\sc iv}$\lambda1548$ is blended with
the C~{\sc iv}$\lambda1550$ at $z_{\rm abs} = 2.1979$. In Fig.~5,
the VPFIT solutions with (red curve) and without (blue curve) taking
into account the contribution of the C~{\sc iv}$\lambda1550$
absorption at $z_{\rm abs} =$2.1979 are over-plotted on the C~{\sc
iv}$\lambda1548$ velocity profile. The high quality of the fit
further confirms the reality of the C~{\sc iv} doublet at $z_{\rm
abs} =$ 2.1979, which was identified solely by its unblended C~{\sc
iv}$\lambda1548$ absorption profile. Furthermore, the Si~{\sc
iv}$\lambda1402$ profile itself is also slightly blended with the
C~{\sc iv}$\lambda1548$ profile at $z_{\rm abs} = 1.8994$, and the
fit  was performed including the contribution of this interloping
C~{\sc iv} absorption. In Fig.~5, the VPFIT solution of this
interloping C~{\sc iv} absorption is over-plotted as a blue curve on
top of the Si~{\sc iv}$\lambda1402$ velocity profile. Figure~5 also
shows velocity profiles of 3 species we chose not to fit (i.e.
C~{\sc iii}, Si~{\sc iii} and O~{\sc vi}). Since the C~{\sc
iii}$\lambda977$ profile is highly saturated, we were unable to
accurately fit it even starting from the C~{\sc ii} profile. The
Si~{\sc iii}$\lambda1206$ is also suffering from blending with some
unidentified forest absorption. The N~{\sc v} doublet is so weak
that only a weak N~{\sc v}$\lambda1238$ absorption could be possibly
present. We regard it as a possible detection. The parameters of the
fit to this N~{\sc v} feature are also listed in Table~5. The O~{\sc
vi} doublet is also weak and appears to be suffering from blends
with forest absorption.

\subsection{$z_{\rm abs}$ = 2.2363 }

This absorption system is established by the presence of the C~{\sc
iv} and O~{\sc vi} doublets. The VPFIT solutions and velocity
profiles are presented in Fig.~5. Table~5 lists the parameters.
Although contaminated with some random forest absorption, the H~{\sc
i} column density is very well constrained at log~$N_{\rm H I} =
14.18 \pm 0.05$ by the steepness of the blue wing of the Ly$\alpha$
absorption profile as well as the presence of the unsaturated
Ly$\delta$ line. Note that the absorption complex visible at the
position of the Ly$\gamma$ absorption of this system is due to the
absorption by Si~{\sc ii}$\lambda1193$ of the system at $z_{\rm abs}
= 1.6359$ (blue curve). Furthermore, due to the blending with some
forest absorption, the fit to the O~{\sc vi} doublet was done
without including the O~{\sc vi}$\lambda1031$ absorption profile.
The VPFIT solutions for the C~{\sc iv} and O~{\sc vi} doublets
required 4 and 6 components, respectively. The two O~{\sc vi}
components at  $v \approx +60$ km~s$^{-1}$ and $v \approx +130$
km~s$^{-1}$ are not clearly seen in the C~{\sc iv} profiles. The two
vertical dotted lines in Fig.~5 indicate the position of the two
strongest components of the O~{\sc vi} doublet profiles.

\section{Physical conditions }

In this section we present detailed analyses of the $z_{\rm abs} =
1.6359$ and $z_{\rm abs} = 1.6720$ sub-DLAs and we construct
photoionization models using CLOUDY (Ferland et al. 1998) version
C10.00 for some interesting systems.

\begin{figure*}
\centering
\includegraphics[bb=54 424 550 683,clip=,width=0.96\hsize]{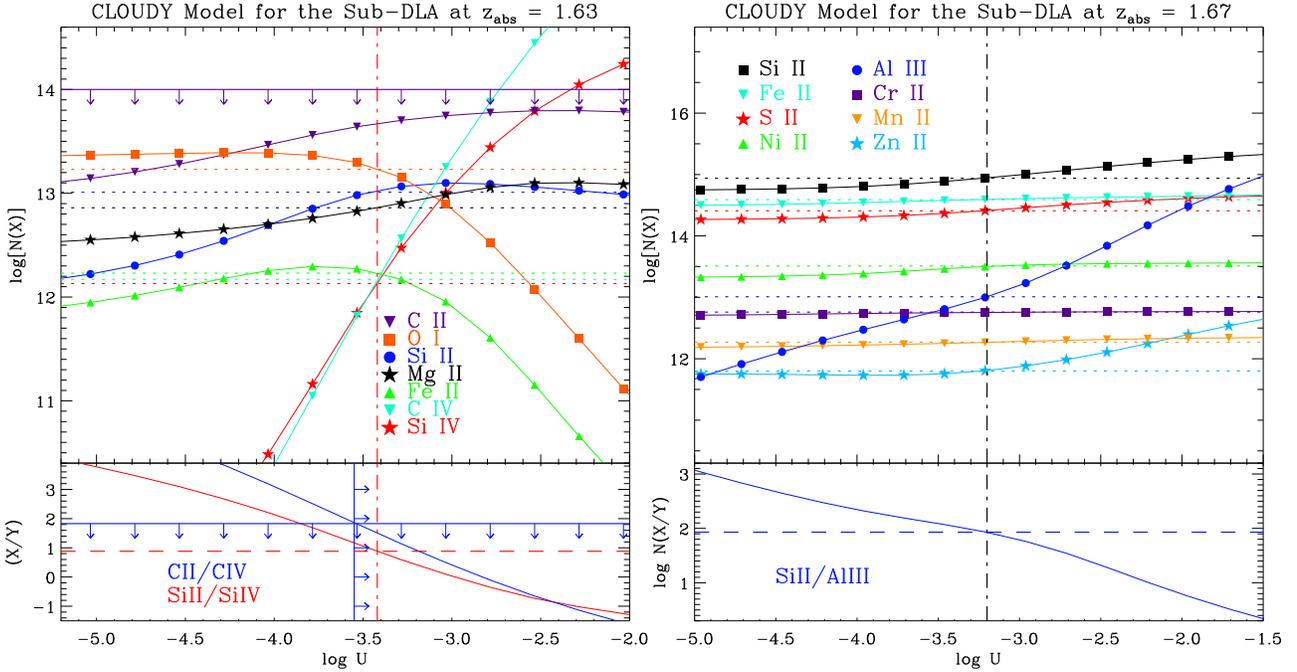}
\caption{CLOUDY models of the two sub-DLAs at $z_{\rm abs} =$ 1.63 and 1.67.}
\end{figure*}

\subsection{Sub-DLA System at $z_{\rm abs} =$ 1.6359}

The best element to derive the overall abundance in the gas is
oxygen, since neutral oxygen and neutral hydrogen have very similar
ionization potentials and are coupled by charge-exchange reaction.
The O~{\sc i} to H~{\sc i} column density ratio yields [O/H] $=
-1.69\pm 0.07$ (corresponding to 1/49 solar) and [O/H] $= -1.65 \pm
0.07$ (corresponding to 1/45 solar) for the R1 and R2 velocity
ranges (see Fig.~1), respectively. This indicates that the whole
complex is well mixed.

A 3~$\sigma$ upper limit on the N~{\sc i} column density implies an
abundance relative to solar [N/H]~$\leq -2.06$ for R2 if we assume
that the ionization fraction is same for nitrogen and hydrogen. The
resulting nitrogen-to-oxygen ratio is [N/O]~$\leq -0.41$ . The
explanation of the N~{\sc i}/H~{\sc i} ratio is not as
straightforward as it is for O~{\sc i}/H~{\sc i}. Although the
ionization potential of hydrogen is slightly lower than that of
neutral nitrogen, ionization effects can considerably affect the
calculation of nitrogen abundance ([N/H]) from N~{\sc i} and H~{\sc
i} in systems with total neutral hydrogen column densities $\leq
10^{20}$ cm$^{-2}$. Thus a non negligible fraction of nitrogen may
be in the form of N~{\sc ii}. While the redshift of this sub-DLA is
too low to observe the corresponding N~{\sc ii} absorption lines
($\lambda_{\rm obs} = 2415, 2585$ \textup{\AA}) from the ground, we
can investigate possible ionization effects in R1 and R2 using other
low ion species, such as Al~{\sc ii} and Al~{\sc iii}. The ratio
Al~{\sc iii}/Al~{\sc ii} is 0.47 and 0.08 in R1 and R2,
respectively, suggesting that the ionization effect is probably
negligible  in the R2 region.

This absorber has an average ratio [Si~{\sc ii}/O~{\sc i}]~$\approx
+0.72$. This ratio is not consistent with nucleosynthesis
considerations, since both Si and O are believed to be produced by
massive stars, and indeed, are observed to have Solar abundance
ratio in Galactic halo stars (Wheeler et al. 1989) and in metal poor
dwarf galaxies (Thuan, Izotov, \& Lipovetsky 1995). One can reconcile
this inconsistency if
some of the Si~{\sc ii}  comes from partially ionized gas rather
than from the neutral gas (recall that the ionization potentials of
O~{\sc i} and Si~{\sc ii} are 13.61, 16.34 eV, respectively).

Since the ionization effects appear to be significant in R1 (judging
from the Al~{\sc iii}/Al~{\sc ii} and [Si~{\sc ii}/O~{\sc i}]
ratios) we try to model this velocity range using model calculations
performed with the photoionization code CLOUDY. The R1 velocity
range comprises five absorption components, and we construct a
detailed model only for the dominant neutral component at $z$ =
1.63588. In this component, the low and high ion species are very
well aligned and we will assume that they arise from the same gas.
The H~{\sc i} column density for this component is estimated to be
log~$N_{\rm H I} = 18.23$ and the calculations were stopped when
this column density was reached. Relative metal abundances are
considered solar and the metallicity is taken as $Z = 0.02
Z_{\odot}$. This metallicity is actually the O~{\sc i} abundance for
the component of interest. We use an ionizing spectrum that is a
combination of the Haardt-Madau (H$\&$M) extragalactic spectrum
(Haardt \& Madau 1996) at $z = $ 1.63, the CMB radiation at $z = $
1.63 and the average Galactic interstellar medium (ISM) spectrum of
Black (1987) with H~{\sc i} ionizing photons extinguished. Note that
we used the CLOUDY built-in HM96 spectrum which includes both
contributions from quasars and galaxies.

In order to determine the ionization parameter, we match the
observed Si~{\sc ii}/Si~{\sc iv} and C~{\sc ii}/C~{\sc iv} ratios to
the values obtained from the model. After the ionization parameter
is derived, the ionization corrections are calculated. The
fractional density of an element in a given ionization state is
defined as $f(X^{i+}) = \frac{N(X^{i+})}{N(X)}$, and similarly for
hydrogen $f(HI) = \frac{n(HI)}{n(H)} = \frac{N(HI)}{N(H)}$. The
ionization corrections IC(X/H) are then given by:

\begin{equation}
IC(X/H) = \log(\frac{N(X^{i+})}{N(HI)}) -log(\frac{N(X)}{N(H)}),
\end{equation}
or
\begin{equation}
IC(X/H) = \log(\frac{f(X^{i+})}{f(HI)}).
\end{equation}

These values are then subtracted from the ionic abundances to obtain
the final ionization-corrected abundance of an element (see
Table~6).

The lower panel of Fig.~6 (left hand side) gives the Si~{\sc
ii}/Si~{\sc iv} and C~{\sc ii}/C~{\sc iv} ratios versus the
ionization parameter log U. The observed ratios intercept the model
curves at the ionization parameter log U = $-$3.42. The upper panel
of Fig.~6 (left hand side) gives the calculated column densities
versus the ionization parameter for different species. In this
figure, the horizontal lines indicate the observed values. The raw
(before correction) and ionization-corrected abundances as well as
the abundances relative to oxygen are reported in Table~6. We note
that hydrogen is 96 percent ionized in this component. The Fe and Si
abundances relative to oxygen ([Fe/O] = $-$0.37$\pm$0.09 and [Si/O]
= $-$0.13$\pm$0.09) are consistent with the mean dust depletion
pattern seen in DLAs and in the Galactic halo (e.g. Petitjean et al.
2002). The Si abundance relative to oxygen also suggests that the
dust depletion in this component is not very significant and that
the underabundance of iron relative to oxygen ([Fe/O] =
$-$0.37$\pm$0.09) could be attributed to nucleosynthetic
considerations. We also note that the calculated $N$(Mg~{\sc
i})/$N$(Mg~{\sc ii}) ratio of $-$2.08 is in good agreement with the
observed value (i.e. $-$2.11). We did not try to model the R2
velocity range because, not only its nine absorption components are
heavily blended, but also none of the components is well aligned in
the high and low ionization species.
 This lack of alignment prevents us from constraining the
ionization parameter using their column density ratios. However, the
observed abundances of Si, Fe and Mg for this velocity range are as
follows: [Si/H] $= -0.99 \pm 0.06$, [Fe/H] $= -1.63 \pm 0.06$ and
[Mg/H] $= -1.54 \pm 0.06$.

 \begin{table}
\begin{minipage}{8.5cm}
\caption{Elemental abundances before ([X/H]$_{raw}$) and after
([X/H]$_{corr}$) ionization correction (IC(X/H)) for the sub-DLA at
$z_{abs} = 1.6359$. These are the results of the CLOUDY model
constructed for the single component at $z_{\rm abs} =$ 1.63588.}
\setlength{\tabcolsep}{8.5pt}
\begin{tabular}{c c c c c}
\hline\hline

 (X) &  [X/H]$_{raw}$  & IC(X/H) & [X/H]$_{corr}$  &  [X/O]                   \\
\hline

O   &                 $-$1.69$\pm$0.07 & $-$0.06   &  $-$1.63$\pm$0.07 &   +0.00                    \\
Si  &                 $-$0.73$\pm$0.06 & +1.03   &  $-$1.76$\pm$0.06 &   $-$0.13                    \\
Fe  &                 $-$1.50$\pm$0.06 & +0.50   &  $-$2.00$\pm$0.06 &   $-$0.37                    \\
Mg  &                 $-$0.97$\pm$0.06 & +0.34   &  $-$1.31$\pm$0.06 &   +0.32                    \\

 \hline
\end{tabular}
\renewcommand{\footnoterule}{}
  \end{minipage}
\end{table}

\begin{figure}
\centering
\includegraphics[bb=121 362 483 717,clip=,width=0.96\hsize]{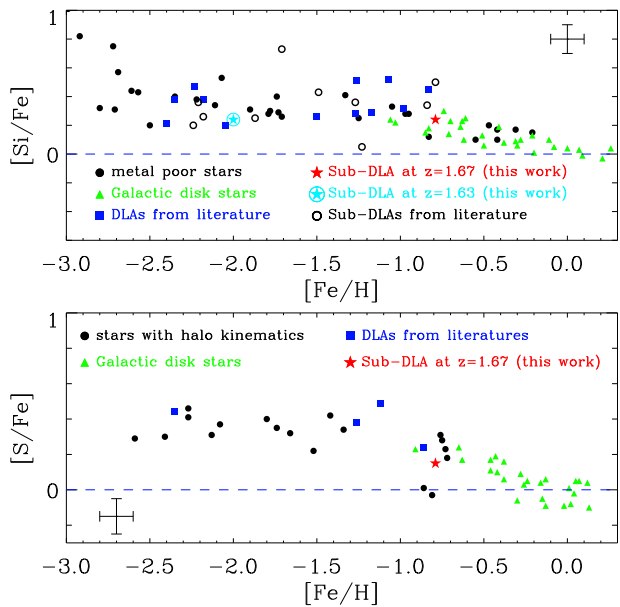}
\caption{[Si/Fe] and [S/Fe] abundance ratios against [Fe/H] as
observed in DLAs and stars in the Galaxy. The values observed in two
sub-DLAs along the studied line of sight are indicated by stars.
The error bars give an indication of the typical uncertainty (for more detail see Sect. 4.2).}
\end{figure}

\subsection{Sub-DLA System at $z_{\rm abs} =$ 1.6720}

In this sub-DLA, the two velocity ranges R2 and R3 exhibit extremely
different ionization properties. In region R2, the observed
$N$(S~{\sc ii})/$N$(O~{\sc i}) ratio is close to unity when the
Solar metallicity of oxygen is 1.56~dex larger than that of sulphur.
Note that similar ratios are observed with Si~{\sc ii}. This
probably means that oxygen, and thus hydrogen, are highly ionized.
Assuming the S/O abundance ratio is solar would imply hydrogen is
ionized at 97\% in this velocity range.If true, this is in
contradiction with the low ionization conditions suggested by the
presence of strong C~{\sc i} absorption. One possibility to escape
this contradiction would be that C~{\sc i} originates in a narrow
and weak component which is lost in wider and stronger O~{\sc i} and
S~{\sc ii} profiles. However, it can be seen on Fig.~2 that this is
not the case and the C~{\sc i}, O~{\sc i}, S~{\sc ii} and Si~{\sc
ii} absorption profiles are similar. This is therefore an intriguing
situation possibly calling for very special abundance ratios.
\par\noindent
In contrast, the R3 velocity range appears to be moderately ionized.
To investigate the ionization effects on the observed abundances, we
constructed a series of CLOUDY photoionization models for the R3
velocity range. We adopted a neutral hydrogen column density of
log~$N_{\rm H I} = 19.78$, a metallicity of $-0.54$ (from the
observed Zn~{\sc ii} and H~{\sc i} column densities), and Solar
relative abundances. As can be seen in the lower panel of Fig.~6
(right hand side), the observed ratio log~$N$(Si~{\sc
ii})/$N$(Al~{\sc iii})~=~1.93 indicates that the ionization
parameter should be close to log U = $-3.20$. Note that due to the
severe saturation of the Al~{\sc ii} absorption profile, Si~{\sc ii}
is used instead of Al~{\sc ii}. The calculated column densities are
given versus ionization parameter in the upper panel of Fig.~6
(right hand side). In this figure, the horizontal dotted lines mark
the observed quantities. Table~7 reports the abundances (before and
after ionization correction), the ionization correction (IC), and
the under/over-abundances relative to Zinc. We note that, in this
velocity range, hydrogen is 50 percent ionized.

The $\alpha$-elements we measure in this sub-DLA follow each other
very well. We obtain [Si/Zn] = +0.02 $\pm$ 0.07 and [S/Zn] = $-$0.07
$\pm$ 0.07, as well as [Si/Fe] = +0.24 $\pm$ 0.07 and [S/Fe] = +0.15
$\pm$ 0.07 after correction of ionization effects. Figure~7 shows
the abundance ratios of Si and S relative to Fe for this sub-DLA
(red stars), along with that of the Galactic and halo stars, DLAs
and sub-DLAs compiled from the literature (see below for
references). The value of [Si/Fe] in this absorber is in good
agreement with measurements of metal poor and Galactic disk stars,
but is relatively low when compared with the sample of DLAs and
sub-DLAs from literature (see upper panel of Fig.~7). The same
behavior is seen in sulphur as [S/Fe] is also lower than the values
in the DLA sample. The [S/Fe] abundance ratio at the corresponding
[Fe/H] is even lower than that of the halo and Galactic disc stars
(with the exception of the two $\alpha$-deficient stars around
[Fe/H]$\approx$$-0.8$ ; see lower panel of Fig.~7). In constructing
the sample of stellar abundances, the following sources are used:
Gratton \& Sneden (1988, 1991) (Fe, Si); Edvardsson et al. (1993)
(Fe, Si); Nissen et al. (2002) (Fe, S); Chen et al. (2002) (Fe, S).
The sample of DLA abundances are from Lu et al. (1996) (Fe, Si, S);
Ledoux et al. (2006) (Fe, S); Meiring et al. (2011) (Fe, S) while
the sample of sub-DLAs is from Dessauges-Zavadsky et al. (2003).

As discussed above, the observed O~{\sc i} column density is 1.56
dex lower than what is expected if [S/O]~=~0. This enables us to
roughly estimate the total oxygen column density to be log~$N$(O) =
15.60. Moreover, for this velocity range, we have measurements for
Si~{\sc ii} (log~$N$(Si~{\sc ii}) = 13.33) and Si~{\sc iv}
(log~$N$(Si~{\sc iv}) = 13.31) as well as a lower limit for Si~{\sc
iii} (log~$N$(Si~{\sc iii}) $\geq$ 14.10). Therefore [Si/O]
($\equiv$ [Si~{\sc ii} + Si~{\sc iii} + Si~{\sc iv}/O]) for R2 is
$\ge$ $-$0.20, which is consistent with the Solar abundance of Si/O
(i.e. [Si/O] = 0.0).

We also tentatively identified a narrow absorption feature in the
red wing of the damped Lyman-$\alpha$ profile of the system at
$z_{\rm abs} = 1.6359$ as N~{\sc i} absorption associated with this
sub-DLA. This feature is shown in Fig.~8 along with a 3-component
VPFIT solution. The VPFIT solution of the intervening damped
Lyman-$\alpha$ profile is also incorporated into the fit. As clearly
depicted in the two magnified panels of Fig.~8, the final solution
could very well match the observation. The redshifts of the
components are fixed to those of the components seen in R3. If our
identification of this N~{\sc i} feature is correct, then the
estimated total column density of log~$N$(N~{\sc i}) =
13.57$\pm$0.03 would imply a mean abundance of [N/H] $= -
2.04\pm0.06$. For an oxygen abundance relative to solar of
$\sim$$-0.5$, this nitrogen metallicity seems small, indeed, below
the secondary relation expected between the metallicities of these
two elements (e.g. Petitjean et al. 2008). This may indicate that we
have underestimated the N~{\sc i} column density because of the
difficulty of the measurement or that the ionization of nitrogen is
higher than that of oxygen.

We observe that Zn is slightly overabundant relative to Ni, Fe, Cr,
and Mn indicating depletion of these elements onto dust. This is
consistent both with the presence of a small amount of ISM-like dust
as well as the abundance pattern of halo stars which have been
enriched by Type II supernovae (SNe). The odd-even effect, namely
the underabundance of odd-Z elements relative to even-Z elements of
the same nucleosynthetic origin, is an observationally established
property of Halo stars. Since Fe is more susceptible to dust
depletion than Mn in the ISM, this ratio can be exploited to
distinguish between dust depletion and pure Type II SNe enrichment.
The [Mn/Fe] = $-$0.24 $\pm$ 0.07 in this sub-DLA, possibly confirms
the odd-even effect, and is consistent with the sample of Halo
metal-poor stars in Ryan et al. (1996). It is interesting to note
that the relative abundance of Mn to Fe in this system, is  in
accordance with that of thick disk stars in Prochaska et al. (2000)
and  what has been observed in DLAs (Ledoux et al. 2002).

The degree of depletion of refractory elements onto dust grains is
in any case small in this sub-DLA. Figure~9 shows the Zn/Fe
abundance ratio against the Zn abundance for this sub-DLA (red
square) and sub-DLAs from the literature (black dots; see below for
references). Figure~9 indicates that the amount of dust depletion in
this sub-DLA is consistent with that seen in other sub-DLAs with the
same metallicity. In constructing the sample of sub-DLA abundances,
the following sources are used: Dessauges-Zavadsky et al. (2003),
P\'eroux et al (2008), Meiring et al. (2007), and Meiring et al.
(2009). Furthermore, the relative abundance ratio of Cr and Fe in R3
is almost solar ([Fe/Cr] $= -0.07 \pm 0.07$). In Fig.~10 we plot the
abundance of Ni, Mn, Cr, Fe and Si relative to Zn as well as Si
relative to Fe. Except for the abrupt change around 55 km s$^{-1}$,
the variation in the relative abundances is of the order of 0.2 dex.

 \begin{table}
\begin{minipage}{8.5cm}
\caption{The same as Table~6 but for the sub-DLA at $z_{\rm abs} =
1.6720$.}
\setlength{\tabcolsep}{8.3pt}
\begin{tabular}{c c c c c}
\hline\hline

 (X) &  [X/H]$_{raw}$  & IC(X/H) & [X/H]$_{corr}$  &  [X/Zn]                   \\
\hline

Zn   &                 $-$0.54$\pm$0.05 & +0.03   &  $-$0.57$\pm$0.05 &   +0.00                    \\
Si   &                 $-$0.35$\pm$0.05 & +0.20   &  $-$0.55$\pm$0.05 &   +0.02                   \\
Fe   &                 $-$0.69$\pm$0.05 & +0.10   &  $-$0.79$\pm$0.05 &   $-$0.22                   \\
Cr   &                 $-$0.66$\pm$0.05 & +0.06   &  $-$0.72$\pm$0.05 &   $-$0.15                   \\
Ni   &                 $-$0.49$\pm$0.05 & +0.19   &  $-$0.68$\pm$0.05 &   $-$0.11                   \\
Mn   &                 $-$0.94$\pm$0.05 & +0.09   &  $-$1.03$\pm$0.05 &   $-$0.46                   \\
S    &                 $-$0.49$\pm$0.05 & +0.15   &  $-$0.64$\pm$0.05 &   $-$0.07                   \\

 \hline
\end{tabular}
\renewcommand{\footnoterule}{}
  \end{minipage}
\end{table}

\begin{figure}
\includegraphics[bb=70 360 558 720,clip=,width=1.01\hsize]{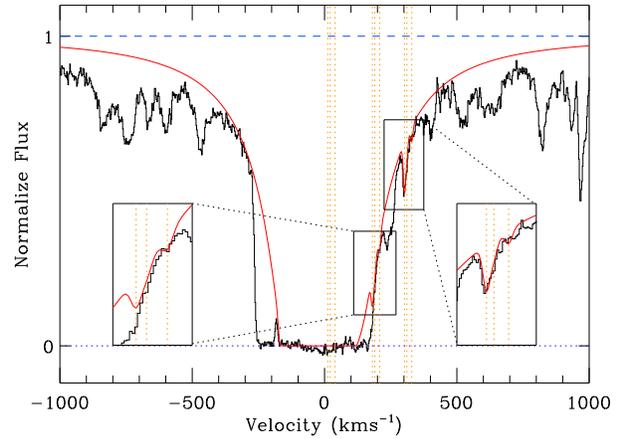}
\caption{N~{\sc i} absorption of the system at $z_{\rm abs} =$
1.6720 in the wing of the Ly $\alpha$ profile of a system at $z_{\rm
abs} =$ 1.6359. The orange vertical lines indicate the position of
the three
components identified in the system for the N~{\sc i} triplet (i.e. $\lambda\lambda\lambda$1200.7,1200.2,1199.5).}
\end{figure}

\begin{figure}
\includegraphics[bb=86 371 543 695,clip=,width=0.96\hsize]{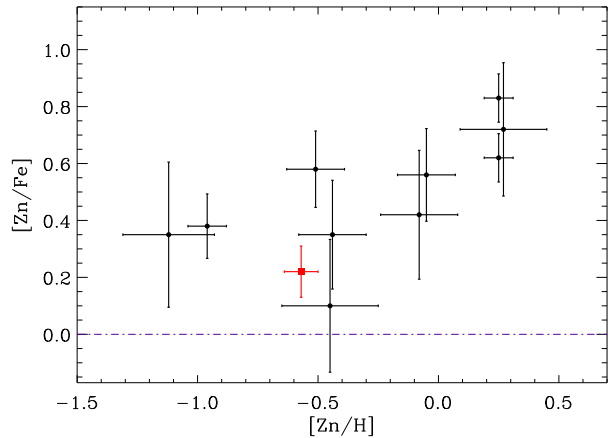}
\caption{ [Zn/Fe] abundance ratio versus [Zn/H] for the sub-DLA at
$z_{\rm abs} =$ 1.6720 (red square) as well as the sub-DLAs from
the literature (black dots; see the text for the references). }
\end{figure}

\begin{figure}
\includegraphics[bb=62 364 332 717,clip=,width=0.90\hsize]{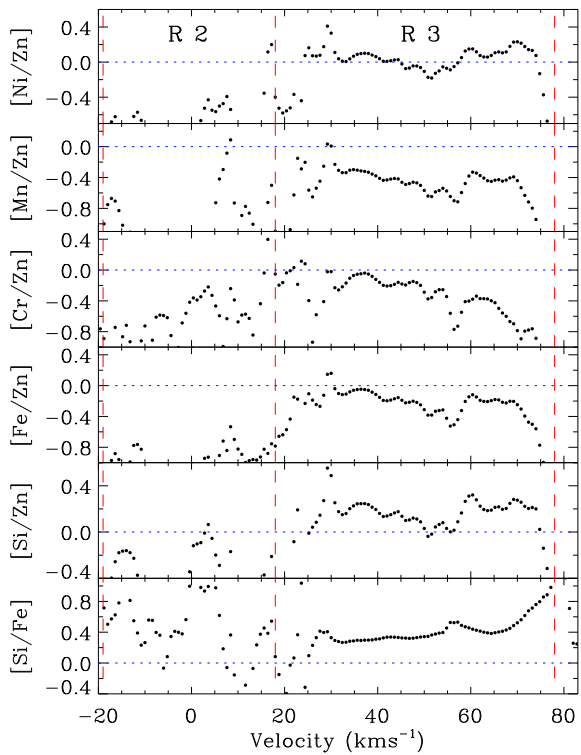}
\caption{Abundances of Ni, Mn, Cr, Fe and Si relative to Zn and
Si relative to Fe in the $z_{\rm abs} =$ 1.6720 sub-DLA. The blue
horizontal dot-dashed line represents zero depletion with respect to Zn
(or with respect to Fe in the case of [Si/Fe]). It can be seen that the
profile is fairly homogeneous.}
\end{figure}

\subsection{Modeling of some systems with $z_{\rm abs} \ge 2.0$}

In this subsection we construct photoionization models using CLOUDY
for the $z_{\rm abs} =$ 2.0422, 2.1979,  2.2028 and 2.2363
absorption systems. In the CLOUDY models of the first two systems we
assumed that the extragalactic UV background of Haardt $\&$ Madau
and the CMB radiation are the radiations striking the absorbing
cloud. For the last two systems, we considered the radiation field
of the active galactic nuclei (AGN) deduced by Mathews \& Ferland
(1987). The relative abundances of the elements were assumed to be
solar (see Table~1).

The reason to model these systems is to test whether these systems
are under the influence of the quasar or not. We will conclude that
it is the case for systems at $z_{\rm abs}$ = 2.0422, 2.2028 and
2.2363 so that these systems will not be considered in the
investigation of the clustering properties of C~{\sc iv} systems
along this line of sight.

\subsubsection{$z_{\rm abs}$ = 2.0422}

As can be seen in Fig.~5, this system comprises two absorption
complexes located at $-$110 $\le v \le$ $-$70 km~s$^{-1}$ and $-$25
$\le v \le$ +40 km~s$^{-1}$. We chose to model the latter complex
for which we have detected a number of high-ion species (i.e. C~{\sc
iv}, N~{\sc v}, and O~{\sc vi}). The CLOUDY model was calculated for
$N$(H~{\sc i}) = 10$^{12.90}$ cm$^{-2}$. The results of the
photoionization model  are illustrated in the upper left panel of
Fig.~11. Using the ionic ratio $N$(O~{\sc vi})/N(C~{\sc iv}) = +0.39
to constrain the ionization parameter gives log~U = $-$1.59. This
ionization parameter along with a metallicity of Z = 6~Z$_{\odot}$
could successfully reproduce the observed C~{\sc iv} and O~{\sc vi}
column densities, but would fail to do so for the N~{\sc v} column
density which appears to be about 0.21 dex larger. Raising by 0.21
dex the nitrogen metallicity we could fit the observation very well.
Note that in Fig.~11 the CLOUDY model of this system is calculated
by incorporating the adjusted metallicities into the code. Given the
high metallicity and ionization parameter, this system is probably
within the sphere of influence of the quasar (Petitjean et al.
1994). We note that the velocity separation between the quasar
($z_{\rm em}$ = 2.233) and the absorber is 18230 km~s$^{-1}$.

\subsubsection{$z_{\rm abs}$ = 2.1979}

For this system, a series of CLOUDY models were run with a typical
H~{\sc i} column density log~$N_{\rm H I} = 15.98$ (the range of
column density derived in this system is 15.80~$<$~log~$N$(H~{\sc
i})~$<$~16.22). The column density ratios of log~$N$(C~{\sc
iv})/$N$(Si~{\sc iv}) = 1.01 and log~$N$(Si~{\sc iii})/$N$(Si~{\sc
iv}) = +0.11 yield ionization parameters of log~U = $-2.12$ and
$-1.99$, respectively, thus in good agreement with each other (see
upper right panel in Fig.~11).  A value of log~U = $-2.05$ is thus
adopted, which along with Z = 0.14 Z$_{\odot}$, could successfully
reproduce the observed column densities of Si~{\sc iii}, Si~{\sc iv}
and C~{\sc iv}.

\begin{figure*}
\centering
\includegraphics[bb=71 362 493 717,clip=,width=0.96\hsize]{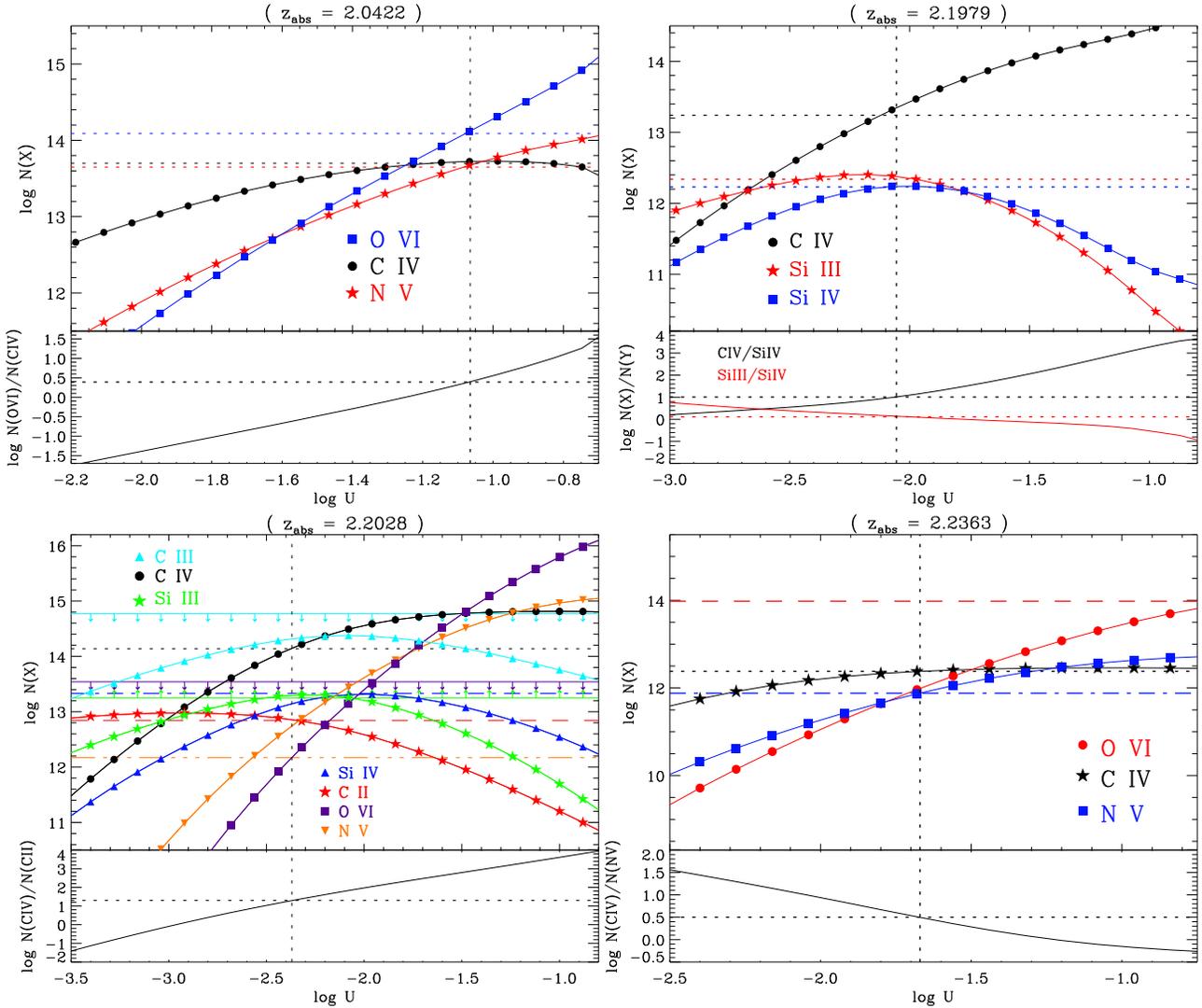}
\caption{CLOUDY photoionization models for the $z_{\rm abs} =$ 2.0422, 2.1979, 2.2028, and 2.2363 absorbers.}
\end{figure*}

\begin{figure*}
\centering
\includegraphics[bb=59 359 554 721,clip=,width=0.96\hsize]{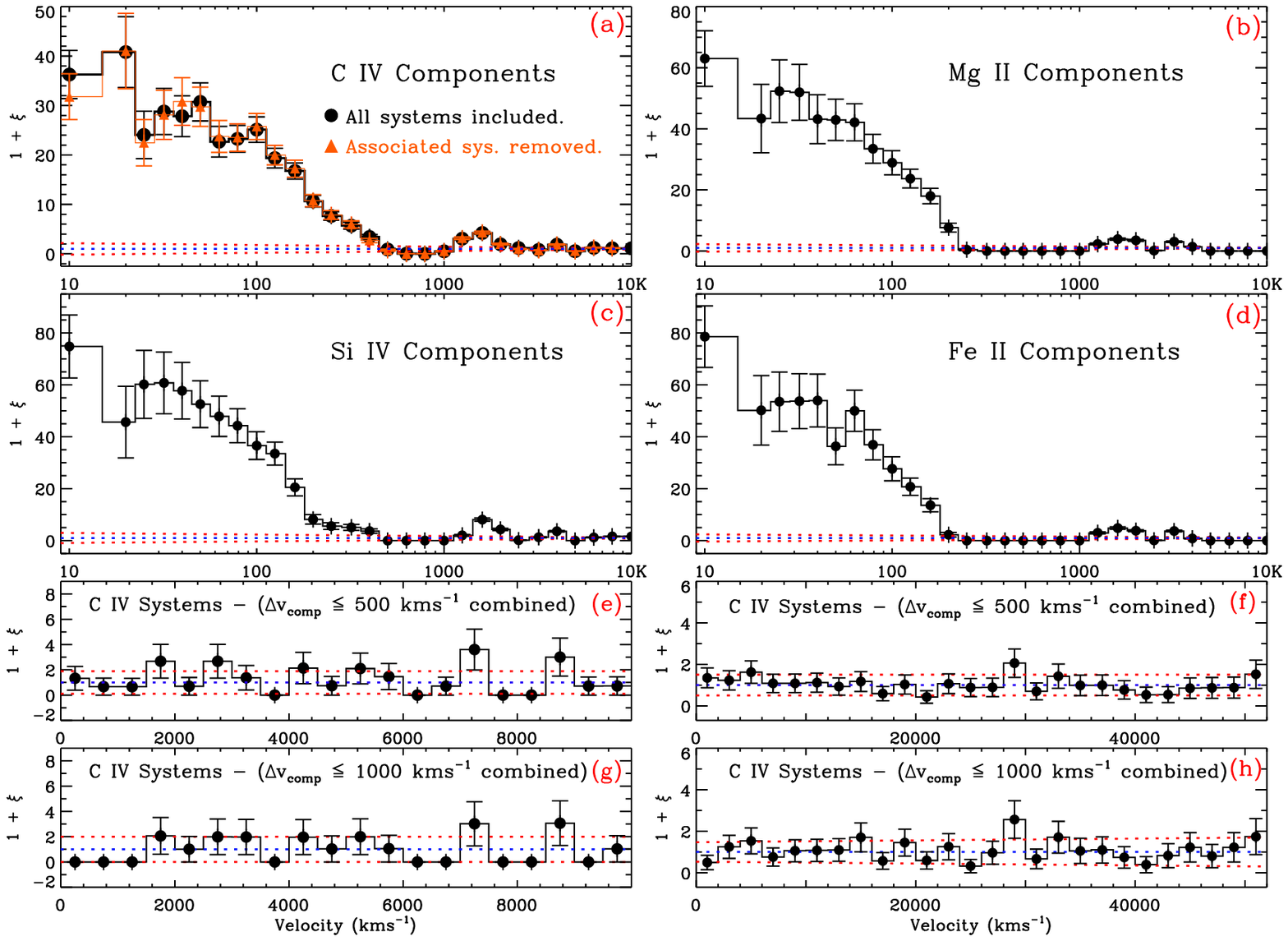}
\caption{  Two-point correlation functions for the C~{\sc iv},
Si~{\sc iv}, Mg~{\sc ii}, and Fe~{\sc ii} absorption components of
all systems detected along this line of sight. The blue-dotted lines
indicate $\xi(v_{k}) = 0$. Panel~(a): The TPCF of the C~{\sc iv}
full sample (note that the orange filled triangles indicate the
C~{\sc iv} TPCF without including the associated systems).
Panel~(b): The TPCF of the Mg~{\sc ii} full sample. Panel~(c): The
TPCF of the Si~{\sc iv} full sample. Panel~(d): The TPCF of the
Fe~{\sc ii} full sample. Panels~(e) \& (f): The TPCF of the C~{\sc
iv} full sample when systems with velocity separation less than 500
km~s$^{-1}$ are combined. Panels~(g) \& (h): The TPCF of the C~{\sc
iv} full sample when systems with velocity separation less than 1000
km~s$^{-1}$ are combined. }

\end{figure*}

\subsubsection{$z_{\rm abs}$ = 2.2028}

The absorption from C~{\sc ii}, C~{\sc iii}, C~{\sc iv}, Si~{\sc
iii}, Si~{\sc iv}, and O~{\sc vi} are spread over about
80~km~s$^{-1}$ in velocity space with very weak associated N~{\sc v}
absorption (see Fig.~5). The grid of CLOUDY models were constructed
for log~$N$(H~{\sc i}) = 15.54. The ionization parameter, log~U =
$-2.37$, was determined using the ionic ratio $N$(C~{\sc
iv})/$N$(C~{\sc ii}). Adopting this ionization parameter, the model
could successfully reproduce the C~{\sc ii} and C~{\sc iv} column
densities with Z = 5.4 Z$_{\odot}$, but failed to do the same for
the Si~{\sc iii}, Si~{\sc iv}, and N~{\sc v}. To reproduce the
observations, the relative abundance of silicon (resp. nitrogen) has
to be raised (resp. lowered) by  0.19 dex (resp. 0.59 dex).
The upper limits to the column densities of O~{\sc vi} and C~{\sc iii} are
also consistent with the model. Note that in Fig.~11 we incorporated
the adjusted metallicities into the CLOUDY model of this system.

\subsubsection{$z_{\rm abs}$ = 2.2363}

This system has a redshift higher than the QSO by 306 km~s$^{-1}$ and
the N~{\sc v} absorption feature is detected albeit very weak. So it
is possible that this system is under the influence of the
ionizing radiation coming from the central engine.

The H~{\sc i} column density is very well constrained, thanks to the
presence of the less blended unsaturated Ly$\delta$ absorption
profile (Fig.~5). As is depicted in Fig.~5, the O~{\sc vi} and
C~{\sc iv} absorption profiles are not well aligned, indicating that
these ions might not arise from exactly the same region and that the
gas could be inhomogeneous. \\

CLOUDY models with log~$N_{\rm H I} = 14.18$ produce an ionic ratio
log~$N$(C~{\sc iv})/$N$(N~{\sc v}) = +0.5 for an ionization
parameter log~U = $-1.67$ and metallicity of Z = 0.56 Z$_{\odot}$
(see lower right panel in Fig.~11). As illustrated in Fig.~11, this
model fails to reproduce the O~{\sc vi} column density by 2.0~dex.
Either relative metallicities are far from solar which could be
explained by the proximity of the gas to the AGN, or this system is
multiphase and the ionization of the O~{\sc vi} phase is much larger
than that of the C~{\sc iv}$-$N~{\sc v} phase. The truth may be
somewhere in the middle.

\section{Clustering properties}

In this section we study the clustering properties of metal lines
(C~{\sc iv}, Si~{\sc iv}, Mg~{\sc ii}, and Fe~{\sc ii}) using the
two-point correlation functions (TPCF) in redshift space, $\xi(v)$.
The velocity two-point correlation function can be calculated from
the pair counts of absorption lines according to :

\begin{equation}
\xi(v_{k}) + 1 = \frac{n_{k}}{<n_{k}>},
\end{equation}
where $n_k$ is the number of observed pairs separated by a velocity
difference $v_{k}$, and $<n_{k}>$ is the average number of such
pairs that would be expected if the systems were randomly
distributed in the absence of clustering. We averaged the output of
1000 Monte Carlo simulations in order to calculate $<n_{k}>$. We
emphasize that we calculate $<n_{k}>$ by simply distributing the
clouds we have found along this line of sight randomly. The
statistical variance in these measurements is given by

\begin{equation}
\sigma^{2} = \frac{n_{k}}{<n_{k}>^{2}}.
\end{equation}

The resulting correlation functions are given in Fig.~12. In this
figure, the blue-dotted lines indicate $\xi(v_{k}) = 0$, and the
1~$\sigma$ region denoted by the two red-dotted lines on both sides
of the $\xi(v_{k}) = 0$ line is determined by the standard deviation
of $\xi(v_{k})$ derived from random simulations. As depicted in
Fig.~12, the overall shapes of the C~{\sc iv} and Si~{\sc iv}
correlation functions are almost similar but the amplitude of the
Si~{\sc iv} TPCF is apparently stronger than that of C~{\sc iv}. In
the case of Mg~{\sc ii} and Fe~{\sc ii}, both the amplitude and the
shape of their TPCFs are similar to that of Si~{\sc iv} to within
their corresponding measurement errors. This is due to the fact that
systems detected only by C~{\sc iv} are spread over large velocity
ranges. In Fig.~12, the C~{\sc iv} TPCF without including the
associated systems is overplotted on the C~{\sc iv} full sample TPCF
as orange filled triangles. As is seen here, at all separations the
TPCF remains practically unchanged.

The velocity correlation length or $v_{0}$, defined as the pair
separation for which $\xi(v_{0}) = 1$, are $v_{0} \approx$ 500
km~s$^{-1}$ for C~{\sc iv} and Si~{\sc iv}, and $v_{0} \approx$ 250
km~s$^{-1}$ for the Mg~{\sc ii} and Fe~{\sc ii} TPCFs. The signal
appears to drop to zero immediately after $v_{0}$ for all four
species contrary to what is seen in Boksenberg et al. (2003, BSR03).
The two-point correlation functions in Fig.~12 exhibit a steep
decline at large velocities ($\ge$ 200 km~s$^{-1}$) and a smoother
decline at small separations, with an elbow occurring at $\approx$
150 km~s$^{-1}$ for the C~{\sc iv} and Si~{\sc iv} profiles and
$\approx$ 100 km~s$^{-1}$ for the Mg~{\sc ii} and Fe~{\sc ii}
profiles. As is seen here and was also noted in Scannapieco et al.
(2006), the elbow occurs at smaller velocity separations for the
Mg~{\sc ii} and Fe~{\sc ii} TPCFs in comparison with that of C~{\sc
iv} and Si~{\sc iv}. Sargent et al. (1988) and Heisler et al. (1989)
found some excess in $\xi_{v}$ between 1000 and 10000 km~s$^{-1}$
and some deficit between 10000 and 20000 km~s$^{-1}$. In our
analysis we detect some excess for the same velocity range but no
deficit is seen for any velocity separation. We tried to further
investigate the clustering signals on large scales by combining all
C~{\sc iv} absorption components spreading less than 500 km~s$^{-1}$
into a single system (panels e \& f in Fig.~12). Here we detect
several peaks in the velocity ranges 1200 -- 3200 km~s$^{-1}$, 4000
-- 5600 km~s$^{-1}$, and 7000 -- 9000 km~s$^{-1}$. There is also a
strong peak in the velocity bin 29000 km~s$^{-1}$.

We considered the possibility that these signals could be artifacts
of the C~{\sc iv} complex decomposition. To this end, we collapsed
complexes with component separations $\Delta v$ $\leq$ 1000
km~s$^{-1}$ into single systems. The results are illustrated in
Fig.~12 panels g and h. The first point to note is that the signals
in the velocity range 1200 $-$ 5600 km~s$^{-1}$ which were
relatively strong in panel e of Fig.~12 are now more consistent with
null clustering, suggesting that the signals might have been due to
the subsplitting of the extended systems. However, the two peaks at
the velocities 7000 $-$ 9000 km~s$^{-1}$ are still present (see
panel g of Fig.~12). Careful examination of the data indicates that
the presence of these two peaks is mainly due to the clustering of
the systems at $z$ $=$ (1.59, 1.65) for the signal at $v$ $=$ 7300
km~s$^{-1}$ and $z$ $=$ (1.59, 1.67) for the $v$ $=$ 8700
km~s$^{-1}$ signal. The spacing between the two systems at redshifts
1.65 and 1.67 is 1660 km~s$^{-1}$, which is consistent with the
separation between the two peaks. The amplitude of the correlation
for the two peaks at $v$ $=$ 7300 km~s$^{-1}$ and $v$ $=$ 8700
km~s$^{-1}$ is estimated to be 3.02 $\pm$ 1.74 and 3.06 $\pm$ 1.77,
respectively. Since these values were calculated for the sample
where velocity separations on scales less than 1000 km~s$^{-1}$ have
been removed it could be possible that these signals represent
powers on supercluster scales. In Fig.~12, panels f and h appear to
show some marginally significant signals at large velocities. The
strongest signal is seen in the $v$ $=$ 29000 km~s$^{-1}$ velocity
bin. The correlation amplitude of this signal is estimated to be
2.56 $\pm$ 0.90. A careful look at the data shows that this signal
is getting most of its power from the coupling of the two systems at
redshifts 1.36 and 1.59.

This study of the clustering properties of C~{\sc iv} absorbers
along the line of sight towards PKS~0237$-$233 has revealed that
C~{\sc iv} components tend to cluster strongly on velocity scales up
to 500 km~s$^{-1}$. This is consistent with what has been derived
from studies of the clustering properties of C~{\sc iv} components
from large samples of intervening C~{\sc iv} systems (Petitjean \&
Bergeron 1994, Boksenberg et al. 2003, Scannapieco et al. 2006,
D'Odorico et al. 2010). This strong clustering signal results from
the combination of relative motions of clouds within a typical
galactic halo as well as clustering between galaxies. Some signal is
also seen at higher velocity separations. However, this signal is
below 2 sigma which means that there is no strong evidence to
support the presence of a high concentration of objects along this
line of sight.

\section{Conclusion}

We have presented a high SNR and high spectral resolution spectrum
of QSO PKS~0237$-$233 and have identified most of the absorption
features redward of the quasar Ly$\alpha$ emission. Three of the 18
identified absorption systems are sub-DLAs with $z_{\rm abs} =$
1.3647, 1.6359, and 1.6720. We have analyzed the latter two systems
in more detail, while the one at $z_{\rm abs} =$ 1.3647, with
metallicity [O/H] $\ge$ $-0.33$, was studied in depth by Srianand et
al. (2007).

CLOUDY models indicate that ionization is higher for the region R1
of the system at $z_{\rm abs} =$ 1.6359, in which hydrogen is more
than 95 percent ionized. In this system, we measured abundances
relative to solar : [O/H] $= -1.63 \pm 0.07$, [Si/H] $= -1.76 \pm
0.06$, [Fe/H] $= -2.00 \pm 0.06$, and [Mg/H] $= -1.31\pm0.06$ for
the R1 velocity range and [O/H] $= -1.65 \pm 0.07$, [Si/H] $= -0.99
\pm 0.06$, [Fe/H] $= -1.63 \pm 0.06$ and [Mg/H] $= -1.54 \pm 0.06$
for the R2 velocity range. The abundance of silicon relative to
oxygen (i.e. [Si/O] $= -0.13 \pm 0.09$) in R1 indicates that dust
depletion in this velocity range is not very significant, and that
[Fe/O] $= -0.37 \pm 0.09$ indicates enrichment by Type II SNe.

The second sub-DLA we studied in detail is located at $z_{\rm abs}
=$ 1.6720. The velocity range of the low ions in this absorber
extends over 260 km~s$^{-1}$, and is divided into three regions, R1,
R2, and R3. We have presented a detailed analysis of the R3 velocity
range for which we have detected most of the low ion species.
Ionization models indicate that hydrogen is 50 percent ionized in
R3, and we measured the following abundances of Zn, Si, and S:
[Zn/H] = $-0.57$$\pm$0.05, [Si/H] = $-0.55$$\pm$0.05, and [S/H] =
$-0.64$$\pm$0.05. We note that hidden component analyses of the
Ni~{\sc ii}, Mn~{\sc ii}, Cr~{\sc ii}, and Al~{\sc iii} transitions
did not reveal any significant hidden saturation or hidden
components, and we expect this to be the case for the other low ion
species. Dust depletion in this system is small ([Fe/Zn] $= -0.22
\pm 0.07$). Finally the region R2 seems to show special ionization
and/or metallicity structure.

We have also studied the clustering properties of metal lines
(C~{\sc iv}, Si~{\sc iv}, Fe~{\sc ii} and Mg~{\sc ii}) for this line
of sight. We found that C~{\sc iv} and Si~{\sc iv} (resp. Fe~{\sc
ii} and Mg~{\sc ii}) trace each other closely and their
corresponding correlation functions $\xi_{v}$ show a steep decline
at large separations ($\ge$ 200 km~s$^{-1}$) and a smoother profile
below $\approx$ 150 km~s$^{-1}$ (resp. $\approx$ 100 km~s$^{-1}$).
Some signals are also detected at larger velocities. These signals
are getting most of their amplitude from the coupling of the systems
at redshifts 1.36, 1.59 and 1.67. Despite detecting an unusually
high number of absorption systems along this line of sight, no
compelling evidence is found to convince us that there is a single
strong overdensity of absorption systems here. However, it would be
good to perform deep imaging in the field, to look for a large
concentration of objects which could be responsible for the presence
of the absorption features seen in the spectrum of the QSO
PKS~0237$-$233.

\section*{Acknowledgments}
R.S. and P.P.J. gratefully acknowledge support from the Indo-French
Centre for the Promotion of Advanced Research (Centre Franco-Indien
pour la promotion de la recherche avance) under Project N.4304-2.
H.F.V. is greatly appreciative of the hospitality of IAP and IUCAA
during his visits to these institutes.

\appendix
\section{Description of the Remaining Absorption Systems}

\begin{figure*}
\centering
\includegraphics[bb=56 368 554 711,clip=,width=0.7\hsize]{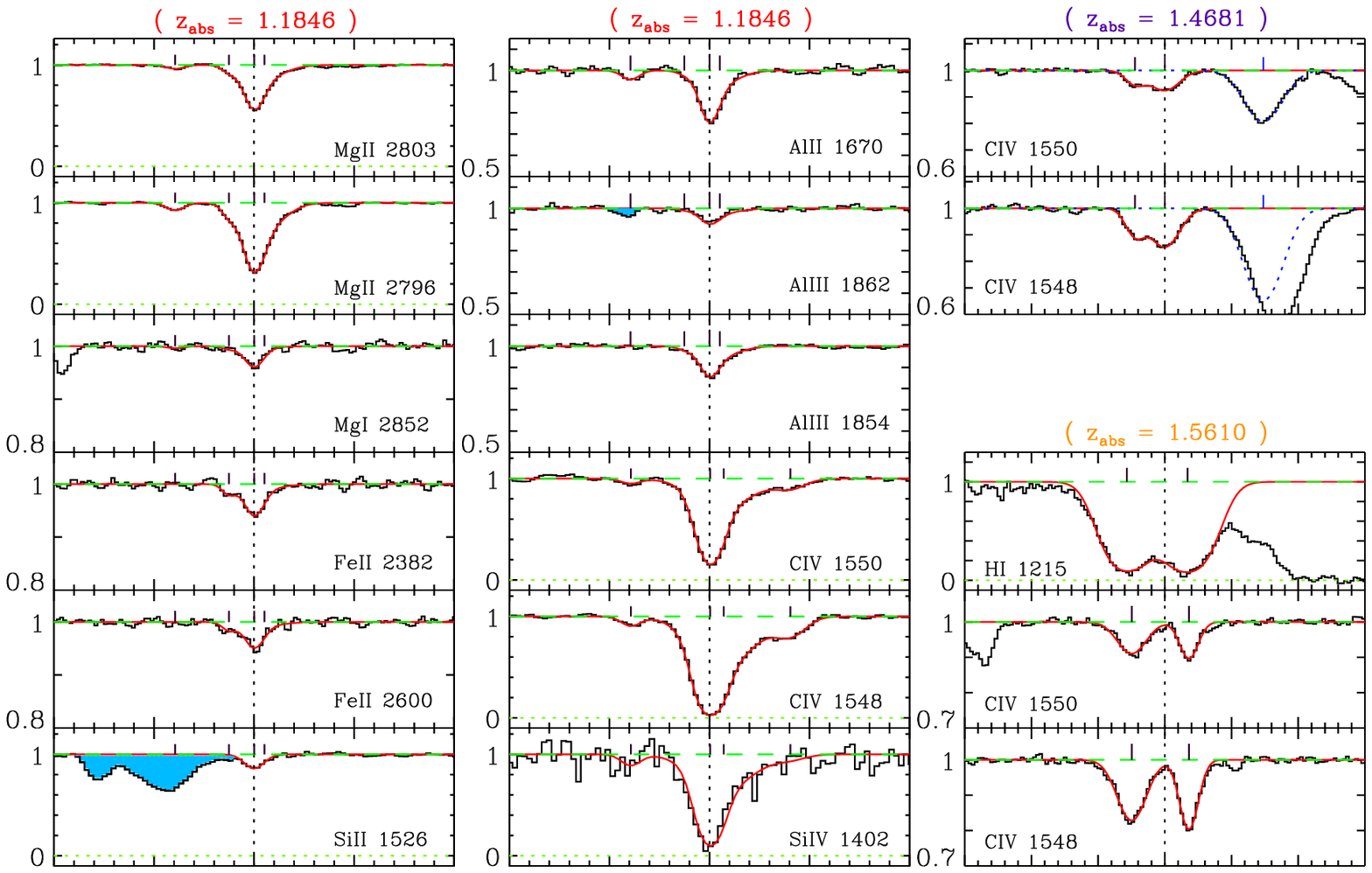}
\caption{Velocity profiles and VPFIT solutions of the absorption
seen in the $z_{\rm abs} =$ 1.1846, $z_{\rm abs} =$ 1.4681, and
$z_{\rm abs} =$ 1.5610 systems. Blue shaded regions indicate blends
with some unrelated features.
Parameters of the fit can be found in Tables~A1 \& A2}
\end{figure*}

\begin{figure*}
\centering
\includegraphics[bb=56 360 459 719,clip=,width=0.7\hsize]{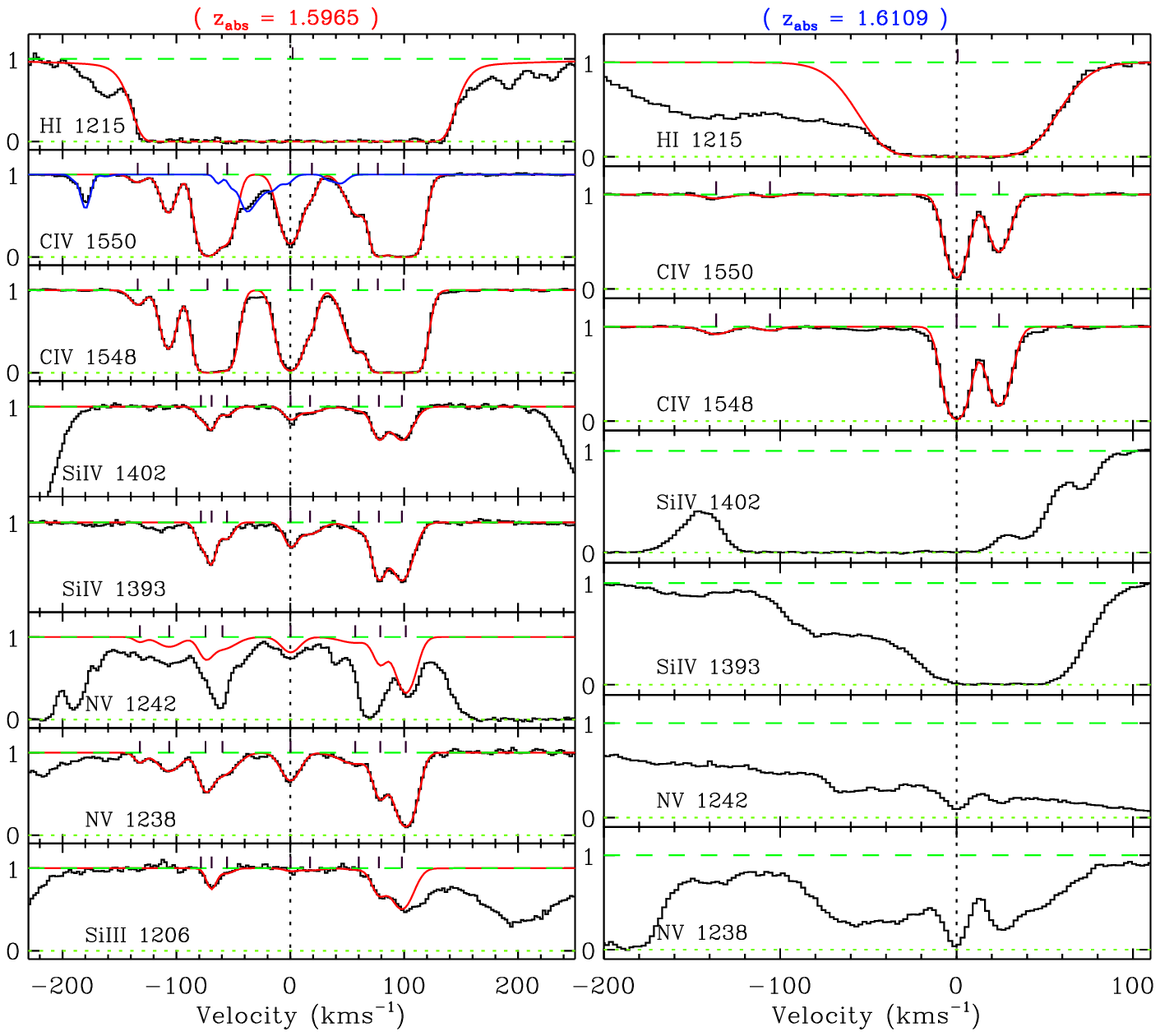}
\caption{Velocity profiles and VPFIT solutions of the absorption
profiles of the systems at $z_{\rm abs} =$ 1.5965, and 1.6109.
Parameters of the fit can be found in Table~A2.}
\end{figure*}

\begin{figure*}
\centering
\includegraphics[bb=53 360 563 721,clip=,width=0.7\hsize]{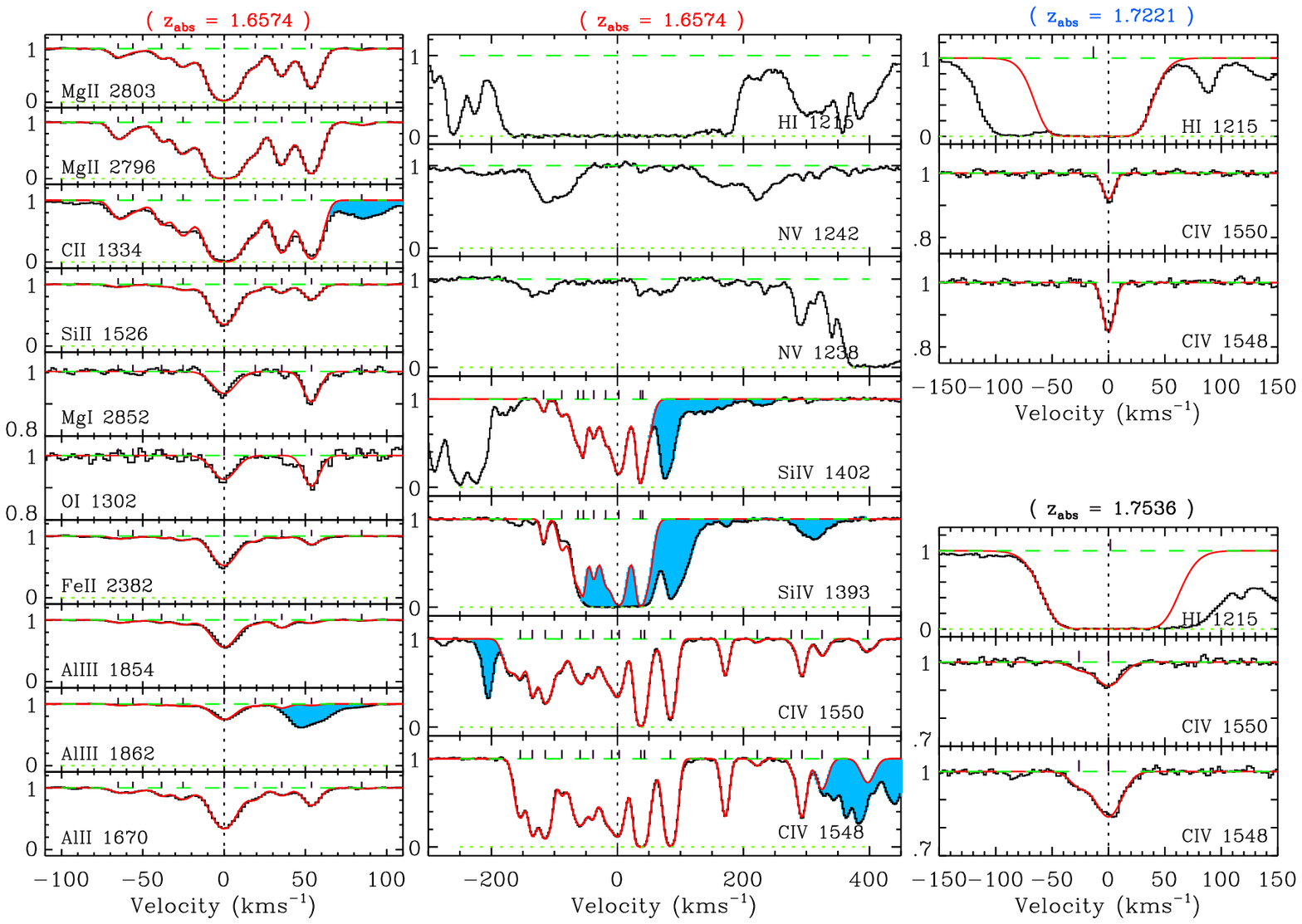}
\caption{Velocity profiles and VPFIT solutions of the low and high
ion transitions in the systems at $z_{\rm abs} =$ 1.6574, 1.7221,
and 1.7536. Blue shaded regions show blends with unrelated
absorption.
Parameters of the fit can be found in Tables~A3 \& A4.}
\end{figure*}

\begin{figure*}
\centering
\includegraphics[bb=64 358 479 720,clip=,width=0.5\hsize]{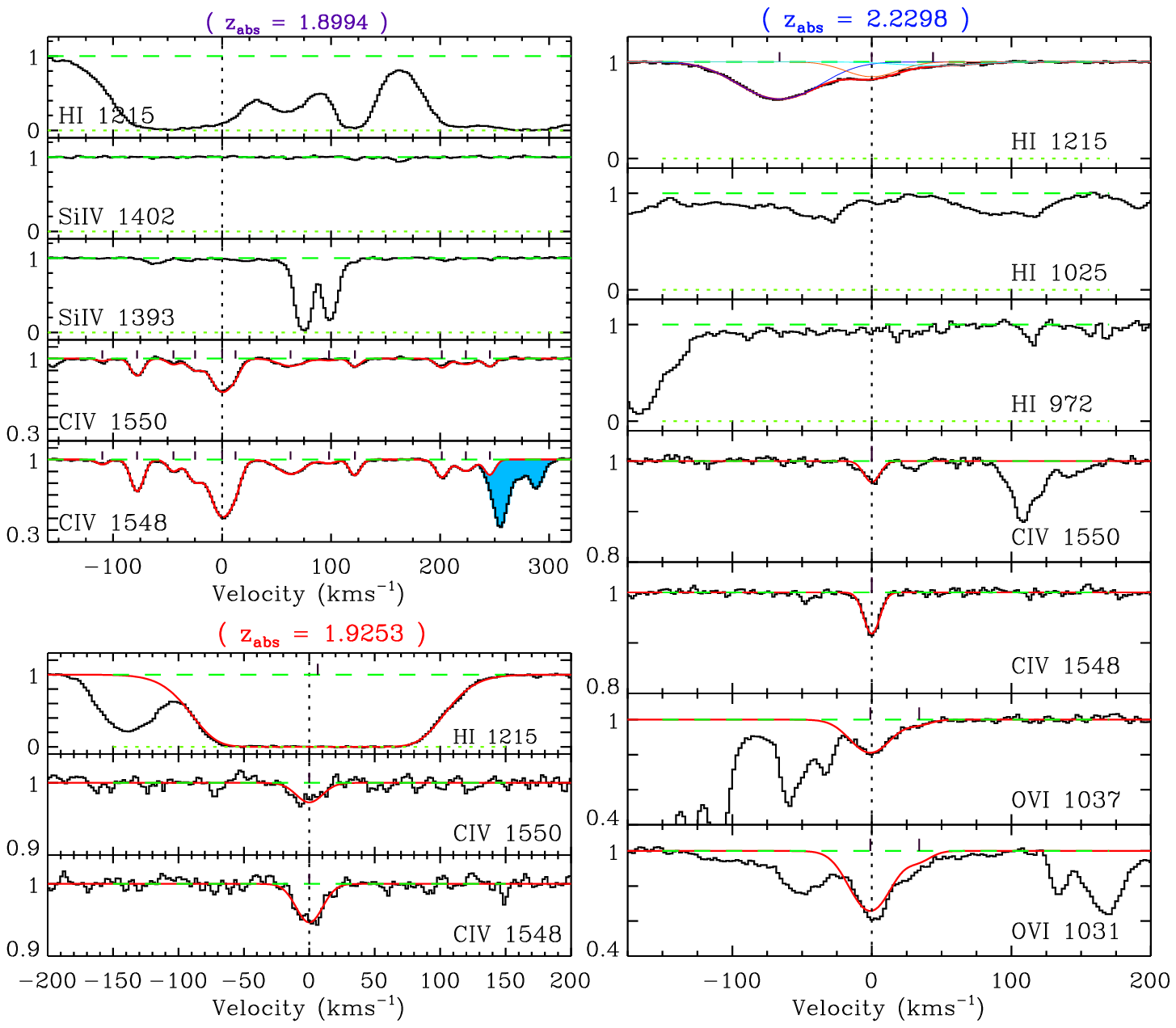}
\caption{Observed velocity profiles and VPFIT solutions of several
transitions seen in the $z_{\rm abs} =$ 1.8994, 1.9253, and 2.2298
absorbers. The blue shaded region shows blends with other lines.
Parameters of the fit can be found in Table~A4.}
\end{figure*}

 \begin{table*}
\begin{minipage}{22.3cm}
\caption{Elemental column densities in the $z_{\rm abs} = 1.1846$
system as well as Solar abundances.}

\setlength{\tabcolsep}{2.0pt}
\renewcommand{\arraystretch}{1.2}
\begin{tabular}{ >{\scriptsize}c >{\scriptsize}c >{\scriptsize}c >{\scriptsize}c >{\scriptsize}c >{\scriptsize}c >{\scriptsize}c >{\scriptsize}c >{\scriptsize}c >{\scriptsize}c
>{\scriptsize}c >{\scriptsize}c}

\hline \hline

\multicolumn{12}{c}{Low-ion column densities} \\
\hline \hline

  $z$    & $\Delta$V (km~s$^{-1}$) & $b$(km~s$^{-1}$) & log$N$(Mg~{\sc ii}) & log$N$(Mg~{\sc i}) & log$N$(O~{\sc i}) & log$N$(Fe~{\sc ii}) & log$N$(Si~{\sc ii}) & log$N$(Al~{\sc ii}) & log$N$(Al~{\sc iii}) & log$N$(C~{\sc ii}) & log$N$(H~{\sc i})\\

\hline

1.18433 &  $-$38.4 &  4.9$\pm$0.4 & 11.27$\pm$0.02 & 9.48$\pm$0.36 & ....\footnote{Not at the range of the data.} & $\leq$10.28 & ....\footnote{Blended with some features.} & 10.80$\pm$0.07 & $\leq$10.65 & ....$^{a}$ & ....$^{a}$ \\
1.18452 &  $-$12.3 &  2.9$\pm$0.7 & 11.35$\pm$0.10 & 9.53$\pm$0.28 & ....$^{a}$ & 10.89$\pm$0.08 & ....$^{b}$ &  10.08$\pm$0.36 & 10.54$\pm$0.27 & ....$^{a}$ & ....$^{a}$\\
1.18461 &   0.0  &  5.6$\pm$0.2 & 12.39$\pm$0.03 & 10.52$\pm$0.03 & ....$^{a}$ & 11.51$\pm$0.04 & 12.48$\pm$0.04 &  11.56$\pm$0.03 & 11.71$\pm$0.04 & ....$^{a}$ & ....$^{a}$\\
1.18465 &  +5.5  & 13.1$\pm$0.8 & 12.16$\pm$0.06 & $\leq$9.50 & ....$^{a}$ & 10.98$\pm$0.18 & 11.83$\pm$0.26 &  11.28$\pm$0.08 & 11.60$\pm$0.08 & ....$^{a}$ & ....$^{a}$\\

\hline \hline

\multicolumn{12}{c}{High-ion column densities} \\
\hline \hline

    &  $z$  & $\Delta$V~(km~s$^{-1}$) &  Ion~(X) & $b$~(km~s$^{-1}$)  &  log~$N$(X)  & $z$  & $\Delta$V~(km~s$^{-1}$) &  Ion~(X) &   $b$~(km~s$^{-1}$)  & log~$N$(X) \\

\hline

    &  1.18433 & $-$38.4 &  C~{\sc iv}    & 5.5$\pm$1.0 & 12.17$\pm$0.05  & 1.18433 & $-$38.4 &  Si~{\sc iv}   & 5.5$\pm$1.0 & 12.16$\pm$0.20 \\
    &  1.18461 &   0.0 &  C~{\sc iv}    & 7.2$\pm$0.5 & 13.79$\pm$0.05   & 1.18462 &   0.0 &  Si~{\sc iv}   & 7.2$\pm$0.5 &  13.50$\pm$0.08 \\
    &  1.18466 &  +6.9 &  C~{\sc iv}    & 19.5$\pm$2.7 & 13.45$\pm$0.07   & 1.18466 &  +6.9 &  Si~{\sc iv}   & 19.5$\pm$2.7 &  13.16$\pm$0.11 \\
    &  1.18491 & +41.2 &  C~{\sc iv}    & 9.8$\pm$2.8 & 12.68$\pm$0.16   & 1.18491 & +41.2 &  Si~{\sc iv}   & 9.8$\pm$2.8 & 11.99$\pm$0.21 \\

\hline

\end{tabular}
\renewcommand{\footnoterule}{}
  \end{minipage}
\end{table*}

\subsection{$z_{\rm abs}$ = 1.1846 }

In addition to the C~{\sc iv} doublet, we identify the Mg~{\sc ii}
and Al~{\sc iii} doublets, Si~{\sc iv}${\lambda}$1402, Si~{\sc
ii}$\lambda1526$, Al~{\sc ii}${\lambda}$1670 and Fe~{\sc ii} lines
associated with this system. Other strong absorption lines,
including Si~{\sc ii}$\lambda1393$ are redshifted outside of our
observed wavelength range. Figure~A1 shows the velocity profiles and
VPFIT solutions of the absorption profiles associated with this
system, and Table~A1 list the parameters (redshift, Doppler
parameter and column density) of the components used to fit the low
and high ion species. The velocity centroids for the components are
indicated by tickmarks above the profile of each transition line. We
found that a minimum of 4 individual components are required to
optimally fit the low ion transitions as well as the high ion
species. Remarkably, there is an extension of the C~{\sc iv} and
Si~{\sc iv} profiles in the red up to about +50~km~s$^{-1}$.

\subsection{$z_{\rm abs} =$ 1.3647 }

This absorber was studied in detail by Srianand et al. (2007) using
the same data. This is a sub-DLA (log~$N_{\rm H I} = 19.30 \pm 0.30$
inferred from IUE data) with near-solar metallicity
([O/H]~$>$~$-0.33$ ). No 21-cm absorption is detected down to
$\tau$(3~$\sigma$)~$<$~3$\times$10$^{-3}$ and the observed C~{\sc i}
excitation indicates that the gas is warm.

\subsection{$z_{\rm abs}$ = 1.4681, 1.5610}

These two absorption systems are very weak and identified solely by
the presence of the C~{\sc iv} and  Ly$\alpha$ absorption (see
Fig.~A1; the Ly$\alpha$ absorption profile of the system at $z_{\rm
abs} =$ 1.4681 is redshifted outside the wavelength range of the
data). Note that there is an absorption feature at $v =$ +70
km~s$^{-1}$ in the $z_{\rm abs} =$1.4681 system which seems to mimic
a C~{\sc iv} doublet. A single component fit was conducted to the
less blended C~{\sc iv} $\lambda$1550 absorption profile. As can be
seen in Fig.~A1 (blue dashed curves), the resulting fit does not
coincide well with the blue wing of the C~{\sc iv}$\lambda$1548
profile, suggesting that the feature may not be a C~{\sc iv} doublet
in the first place. Results of the fits are presented in Table~A2.

 \begin{table*}
 \centering
\caption{High-ion and H~{\sc i} column densities in the $z_{\rm abs}
= 1.4681, 1.5610, 1.5965, 1.6109$ absorption systems.}
 \setlength{\tabcolsep}{4.7pt}
\renewcommand{\arraystretch}{1.2}

\begin{tabular}{c c c c c}

\hline\hline

    $z$ & $\Delta$V~(km~s$^{-1}$)  &  Ion~(X) & $b$~(km~s$^{-1}$) & log~$N$(X) \\

\hline\hline
\multicolumn{5}{c}{$z_{\rm abs} = 1.4681$} \\
\hline
1.46797 & $-$21.9 &  C~{\sc iv}    & 9.2$\pm$0.5  & 12.37$\pm$0.03\\
1.46815 &     0.0 &  C~{\sc iv}    & 14.6$\pm$0.6  & 12.73$\pm$0.02\\
\hline
\multicolumn{5}{c}{$z_{\rm abs} = 1.5610$} \\
\hline
1.56076 & $-$28.1 &  H~{\sc i}     & 20.6$\pm$0.8 & 13.81$\pm$0.02\\
1.56114 & +16.4   &  H~{\sc i}     & 21.5$\pm$1.0 & 13.85$\pm$0.02\\
1.56079 & $-$24.6 &  C~{\sc iv}    & 14.5$\pm$0.3 & 12.81$\pm$0.01\\
1.56115 &  +17.6   &  C~{\sc iv}    & 8.9$\pm$0.2 & 12.68$\pm$0.01\\
\hline
\multicolumn{5}{c}{$z_{\rm abs} = 1.5965$} \\
\hline
1.59655  & +3.5 &      H~{\sc i}      & 46.0   & $\ge$ 17.85 \\
1.59536  & $-$134.0 &  C~{\sc iv}     & 8.3$\pm$1.3   & 12.61$\pm$0.05\\
1.59560  & $-$106.2 &  C~{\sc iv}     & 8.0$\pm$0.2   & 13.40$\pm$0.01\\
1.59590  &  $-$71.6 &  C~{\sc iv}     & 9.2$\pm$0.2   & 14.31$\pm$0.02\\
1.59605  &  $-$54.3 &  C~{\sc iv}     & 8.3$\pm$0.4   & 13.83$\pm$0.03\\
1.59652  &      0.0 &  C~{\sc iv}     & 10.6$\pm$0.2   & 13.97$\pm$0.01\\
1.59669  &  +19.6 &  C~{\sc iv}     & 5.7$\pm$0.8   & 12.89$\pm$0.05\\
1.59704  &  +60.0 &  C~{\sc iv}     & 13.0$\pm$0.7   & 13.63$\pm$0.03\\
1.59719  &  +77.3 &  C~{\sc iv}     & 5.3$\pm$0.5   & 14.29$\pm$0.10\\
1.59739  & +100.4 &  C~{\sc iv}     & 12.2$\pm$0.3   & 14.75$\pm$0.03\\
1.59585  & $-$77.4 &  Si~{\sc iv}     & 5.3$\pm$1.3   & 12.05$\pm$0.20\\
1.59593  & $-$68.1 &  Si~{\sc iv}     & 4.7$\pm$1.5   & 12.32$\pm$0.13\\
1.59604  & $-$55.4 &  Si~{\sc iv}     & 6.8$\pm$1.5   & 12.01$\pm$0.07\\
1.59652  &  0.0 &  Si~{\sc iv}     & 7.2$\pm$0.7   & 12.17$\pm$0.07\\
1.59667  & +17.3 &  Si~{\sc iv}     & 12.8$\pm$3.1   & 12.08$\pm$0.10\\
1.59704  & +60.0 &  Si~{\sc iv}     & 12.4$\pm$4.0   & 11.90$\pm$0.16\\
1.59720  & +78.5 &  Si~{\sc iv}     & 7.5$\pm$0.6   & 12.63$\pm$0.04\\
1.59737  & +98.1 &  Si~{\sc iv}     & 12.2$\pm$0.4   & 12.86$\pm$0.01\\
1.59538  & $-$131.6 &  N~{\sc v}     & 5.2$\pm$1.4   & 12.44$\pm$0.07\\
1.59560  & $-$106.2 &  N~{\sc v}     & 13.1$\pm$1.1   & 13.08$\pm$0.02\\
1.59588  &  $-$73.9 &  N~{\sc v}     & 7.4$\pm$1.0   & 13.17$\pm$0.12\\
1.59601  &  $-$58.9 &  N~{\sc v}     & 15.5$\pm$2.9   & 13.25$\pm$0.11\\
1.59652  &  0.0 & N~{\sc v}     & 10.5$\pm$0.4   & 13.20$\pm$0.01\\
1.59702  & +57.7 &  N~{\sc v}     & 17.7$\pm$3.1   & 12.96$\pm$0.07\\
1.59721  & +79.7 &  N~{\sc v}     & 7.2$\pm$0.5   & 13.35$\pm$0.03\\
1.59740  & +101.6 &  N~{\sc v}     & 10.3$\pm$0.2   & 13.94$\pm$0.01\\
\hline
\multicolumn{5}{c}{$z_{\rm abs} = 1.6109$} \\
\hline
1.61086 & 0.0 &  H~{\sc i}     & 37.1 & $\ge$ 14.59\\
1.60967 & $-$136.7 &  C~{\sc iv}    & 9.5$\pm$1.6 & 12.28$\pm$0.05\\
1.60994 & $-$105.7 & C~{\sc iv}    & 6.7$\pm$2.9 & 11.87$\pm$0.10\\
1.61086 &  0.0 &  C~{\sc iv}    & 6.5$\pm$0.1 & 13.89$\pm$0.01\\
1.61107 & +24.1 &  C~{\sc iv}    & 7.1$\pm$0.1 & 13.54$\pm$0.01\\
1.61086 & 0.0 &  N~{\sc v}     & 6.5$\pm$0.0 & 14.00$\pm$0.03\\
\hline

\end{tabular}
\end{table*}

 \begin{table*}
\begin{minipage}{22.3cm}
\caption{Elemental column densities in the $z_{\rm abs} = 1.6574$
system.}

\setlength{\tabcolsep}{2.0pt}
\renewcommand{\arraystretch}{1.2}
\begin{tabular}{ >{\scriptsize}c >{\scriptsize}c >{\scriptsize}c >{\scriptsize}c >{\scriptsize}c >{\scriptsize}c >{\scriptsize}c >{\scriptsize}c >{\scriptsize}c >{\scriptsize}c
>{\scriptsize}c >{\scriptsize}c}

\hline \hline

\multicolumn{12}{c}{Low-ion column densities} \\
\hline \hline

  $z$    & $\Delta$V (km~s$^{-1}$) & $b$(km~s$^{-1}$) & log$N$(Mg~{\sc ii}) & log$N$(Mg~{\sc i}) & log$N$(O~{\sc i}) & log$N$(Fe~{\sc ii}) & log$N$(Si~{\sc ii}) & log$N$(Al~{\sc iii}) & log$N$(Al~{\sc ii}) & log$N$(C~{\sc ii}) & log$N$(H~{\sc i})\\

\hline

1.65683 &  $-$64.3 &   5.1$\pm$0.9  & 11.87$\pm$0.01 &  $\leq$9.67       &  $\leq$11.89    & 11.32$\pm$0.12 & 11.96$\pm$0.14  & 11.16$\pm$0.10  & 10.86$\pm$0.12   & 12.82$\pm$0.04    & ....\footnote{Too uncertain.}\\
1.65691 &  $-$55.3 &   9.8$\pm$5.7  & 11.68$\pm$0.02 &  $\leq$10.12      &  $\leq$12.39    & 11.25$\pm$0.18 & 11.58$\pm$0.43  & 11.17$\pm$0.12  & 11.20$\pm$0.07   & 12.90$\pm$0.04    & ....$^{a}$\\
1.65706 &  $-$38.4 &   3.5$\pm$0.4  & 11.88$\pm$0.01 &  $\leq$9.66       &  $\leq$11.89    & 11.10$\pm$0.15 & 11.79$\pm$0.16  & 11.13$\pm$0.08  & 11.21$\pm$0.04   & 12.98$\pm$0.03    & ....$^{a}$\\
1.65718 &  $-$24.8 &   6.5$\pm$0.2  & 12.38$\pm$0.00 &  $\leq$9.93       &  $\leq$12.14    & 11.66$\pm$0.05 & 12.40$\pm$0.05  & 11.59$\pm$0.03  & 11.56$\pm$0.02   & 13.40$\pm$0.01    & ....$^{a}$\\
1.65740 &    0.0 &   9.3$\pm$0.1  & 13.45$\pm$0.00 &  10.96$\pm$0.04   &  12.92$\pm$0.09 & 12.76$\pm$0.01 & 13.55$\pm$0.01  & 12.57$\pm$0.00    & 12.40$\pm$0.01   & 14.38$\pm$0.01    & ....$^{a}$\\
1.65757 &  +19.2 &   4.8$\pm$0.3  & 12.20$\pm$0.01 &  $\leq$9.89       &  $\leq$11.67    & 11.59$\pm$0.05 & 12.28$\pm$0.06  & 11.36$\pm$0.05    & 11.45$\pm$0.03   & 13.22$\pm$0.02    & ....$^{a}$\\
1.65772 &  +36.1 &   5.0$\pm$0.1  & 12.55$\pm$0.00 &  $\leq$10.10      &  $\leq$12.28    & 11.25$\pm$0.11 & 12.50$\pm$0.04  & 11.75$\pm$0.02    & 11.30$\pm$0.04   & 13.72$\pm$0.01    & ....$^{a}$\\
1.65788 &  +54.1 &   5.7$\pm$0.0  & 12.85$\pm$0.00 &  10.94$\pm$0.03   &  12.87$\pm$0.08 & 11.94$\pm$0.03 & 12.81$\pm$0.02  & 11.37$\pm$0.05    & 11.73$\pm$0.02   & 13.96$\pm$0.01    & ....$^{a}$\\
1.65815 &  +84.6 &   4.8$\pm$1.3  & 11.08$\pm$0.06 &  $\leq$9.70       &  $\leq$12.02    &    $\leq$10.28 &   $\leq$11.33   & 10.67$\pm$0.24    & 10.26$\pm$0.38   &    ....\footnote{Blended with some features.}     & ....$^{a}$\\

\hline \hline

\multicolumn{12}{c}{High-ion column densities} \\
\hline \hline

    &  $z$  & $\Delta$V~(km~s$^{-1}$) &  Ion~(X) & $b$~(km~s$^{-1}$)  &  log~$N$(X)  & $z$  & $\Delta$V~(km~s$^{-1}$) &  Ion~(X) &   $b$~(km~s$^{-1}$)  & log~$N$(X) \\

\hline
    &  1.65604  & $-$153.5 &  C~{\sc iv}     & 10.6$\pm$0.5   & 13.45$\pm$0.02     & 1.65985  & +276.3 &  C~{\sc iv}     &  8.1$\pm$3.9   &  12.26$\pm$0.13 \\
    &  1.65621  & $-$134.3 &  C~{\sc iv}     & 6.9$\pm$0.5   &  13.54$\pm$0.03     & 1.66000  & +293.2 &  C~{\sc iv}     &  8.5$\pm$0.2   &  13.36$\pm$0.01 \\
    &  1.65639  & $-$114.0 &  C~{\sc iv}     & 12.1$\pm$0.4   &  13.83$\pm$0.01    & 1.66029  & +325.8 &  C~{\sc iv}     &  12.2$\pm$0.3   &  13.11$\pm$0.01 \\
    &  1.65662  & $-$88.0  &  C~{\sc iv}     &  6.4$\pm$0.3   &  12.84$\pm$0.02    & 1.66093  & +398.0 &  C~{\sc iv}     &  15.2$\pm$0.4   &  13.06$\pm$0.01 \\
    &  1.65688  & $-$58.7 &  C~{\sc iv}     &  17.0$\pm$0.4   &  13.75$\pm$0.01    & 1.65636  & $-$117.3 &  Si~{\sc iv}     &  5.4$\pm$1.0   &  12.29$\pm$0.04 \\
    &  1.65706  & $-$38.4 &  C~{\sc iv}     &  5.2$\pm$0.4   &  12.93$\pm$0.04     & 1.65662  & $-$88.0 &  Si~{\sc iv}     &  7.7$\pm$0.9   &  12.52$\pm$0.04 \\
    &  1.65732  &  $-$9.0 &  C~{\sc iv}     &  21.3$\pm$1.7   &  13.87$\pm$0.07    & 1.65687  & $-$59.8 &  Si~{\sc iv}     &  12.0$\pm$1.2   &  13.21$\pm$0.06 \\
    &  1.65743  & +3.4 &  C~{\sc iv}     &  9.3$\pm$1.2   &  13.35$\pm$0.16        & 1.65693  & $-$53.0 &  Si~{\sc iv}     &  4.4$\pm$0.8   &  12.84$\pm$0.11 \\
    &  1.65773  & +37.2 &  C~{\sc iv}     &  8.1$\pm$0.2   &  14.26$\pm$0.02       & 1.65707  & $-$37.2 &  Si~{\sc iv}     &  7.2$\pm$1.1   &  12.93$\pm$0.06 \\
    &  1.65778  & +42.9 &   C~{\sc iv}     &  15.9$\pm$2.2   &  13.56$\pm$0.13     & 1.65724  & $-$18.0 &  Si~{\sc iv}     &  8.7$\pm$1.4   &  13.04$\pm$0.07 \\
    &  1.65815  & +84.6 &  C~{\sc iv}     &  9.5$\pm$0.1   &  14.04$\pm$0.00       & 1.65742  & +2.2 &  Si~{\sc iv}     &  11.7$\pm$0.4   &  13.65$\pm$0.01 \\
    &  1.65892  & +171.4 &  C~{\sc iv}     &  7.1$\pm$0.1   &  13.29$\pm$0.00      & 1.65773  & +37.2 &  Si~{\sc iv}     &  4.2$\pm$0.4   &  13.45$\pm$0.04 \\
    &  1.65937  & +222.2 &  C~{\sc iv}     &  8.6$\pm$0.9   &  12.26$\pm$0.03      & 1.65776  & +40.6 &  Si~{\sc iv}     &  13.8$\pm$1.1   &  13.48$\pm$0.02 \\

\hline

\end{tabular}
\renewcommand{\footnoterule}{}
  \end{minipage}
\end{table*}

\subsection{$z_{\rm abs}$ = 1.5965 }

Figure~A2 shows the velocity profiles and VPFIT solutions of the
neutral hydrogen and metal line transitions associated with this
system. Using the Ly$\alpha$ absorption profile, we could determine
a lower limit to the H~{\sc i} column density of log~$N$(H~{\sc
i})~$\ge$~17.85. We identify metal absorption from C~{\sc iv},
Si~{\sc iv}, N~{\sc v} as well as Si~{\sc iii}${\lambda}$1206, with
no trace of low ion species. The velocity spread is $\approx$300
km~s$^{-1}$. Results of the fits are presented in Table~A2. We chose
not to fit the N~{\sc v}$\lambda1242$ because its profile appears to
be contaminated by some forest absorption. Moreover, the C~{\sc
iv}$\lambda1550$ profile is blended with the Si~{\sc
ii}$\lambda1526$ absorption from the sub-DLA at $z_{\rm abs} =
1.6359$. The VPFIT solution for this Si~{\sc ii} absorption derived
from Si~{\sc ii}$\lambda$1304 alone is over-plotted on top of the
C~{\sc iv}$\lambda1550$ absorption profile as a blue curve in
Fig.~A2. As can be seen in the figure, the fit is not perfect at
velocities of $v\approx -180$, $-50 \leq v \leq -20$ and $v \approx
+35$ km~s$^{-1}$, but clearly within errors.


\subsection{$z_{\rm abs}$ = 1.6109 }

Figure~A2 shows the velocity profiles  of the H~{\sc i}, C~{\sc iv},
Si~{\sc iv} and N~{\sc v} doublets associated with this system. The
Ly$\alpha$ absorption profile of this system gives a lower limit to
the H~{\sc i} column density of log~$N$(H~{\sc i}) $\ge$ 14.59. We
identify no low ion transition for this system and based on the
absence of any Si~{\sc ii} and C~{\sc ii} absorption we conclude
that this is a high ionization system. The N~{\sc v} and Si~{\sc iv}
doublets are heavily blended in the forest, therefore we only
present in Fig.~A2 the VPFIT solution for the C~{\sc iv} doublet for
which we found a 4-component fit was optimal. In addition, we have
also conducted a single component fit to the N~{\sc v} absorption
profile at $v = 0$ km~s$^{-1}$. The parameters are listed in
Table~A2. The two C~{\sc iv} components at $v = -135$ km~s$^{-1}$
and $v = -105~$ km~s$^{-1}$ are very weak but they are certainly
real. Also the O~{\sc vi} doublet for this system is outside of our
observed wavelength range.


 \begin{table*}
 \centering
\caption{High-ion and H~{\sc i} column densities for the $z_{\rm
abs} =$ 1.7221, 1.7536, 1.8994, 1.9253, and 2.2298 absorbers.}
 \setlength{\tabcolsep}{6.1pt}
\renewcommand{\arraystretch}{1.2}

\begin{tabular}{c c c c c}

\hline\hline

    $z$ & $\Delta$V~(km~s$^{-1}$)  &  Ion~(X) & $b$~(km~s$^{-1}$) & log~$N$(X) \\

\hline
\multicolumn{5}{c}{$z_{\rm abs} = 1.7221$} \\
\hline
1.72198 & $-$14.3 &  H~{\sc i}     & 30.3 & $\ge$ 14.80\\
1.72211 & 0.0 &  C~{\sc iv}    & 6.0$\pm$0.2 & 12.43$\pm$0.01\\
\hline
\multicolumn{5}{c}{$z_{\rm abs} = 1.7536$} \\
\hline
1.75357 & +1.1 &  H~{\sc i}     & 36.1 & $\ge$ 14.81\\
1.75332 & $-$26.1 &  C~{\sc iv}    & 11.3$\pm$1.4 & 12.13$\pm$0.06\\
1.75356 & 0.0 &  C~{\sc iv}    & 15.5$\pm$0.6 & 12.80$\pm$0.01\\
\hline
\multicolumn{5}{c}{$z_{\rm abs} = 1.8994$} \\
\hline
1.89838  & $-$110.6 &   C~{\sc iv} &   6.3$\pm$1.0     &   11.83$\pm$0.04\\
1.89869  & $-$78.6 &   C~{\sc iv} &   7.1$\pm$0.1     &   12.76$\pm$0.01  \\
1.89901  & $-$45.5 &   C~{\sc iv} &   6.8$\pm$0.6     &   12.26$\pm$0.03  \\
1.89921  & $-$24.8 &   C~{\sc iv} &   8.7$\pm$0.6     &   12.63$\pm$0.04  \\
1.89945  & 0.0 &   C~{\sc iv} &   12.7$\pm$0.9     &   13.30$\pm$0.05  \\
1.89957  & +12.4 &   C~{\sc iv} &   6.6$\pm$2.3     &   12.35$\pm$0.32  \\
1.90005  & +62.0 &   C~{\sc iv} &   15.8$\pm$0.5     &   12.68$\pm$0.01  \\
1.90039  & +97.2 &   C~{\sc iv} &   8.2$\pm$1.6     &   11.81$\pm$0.06  \\
1.90062  & +120.9 &    C~{\sc iv} &   5.9$\pm$0.3     &   12.36$\pm$0.01  \\
1.90140  & +201.6 &   C~{\sc iv} &   6.8$\pm$0.3     &   12.46$\pm$0.02  \\
1.90161  & +223.3 &  C~{\sc iv} &   9.8$\pm$0.9     &   12.39$\pm$0.03  \\
1.90182  & +244.9 &  C~{\sc iv} &   5.7$\pm$0.8     &   12.33$\pm$0.04  \\
\hline
\multicolumn{5}{c}{$z_{\rm abs} = 1.9253$} \\
\hline
1.92533 & 0.0 &  H~{\sc i}     & 54.6 & $\ge$ 15.12\\
1.92527 & 0.0 & C~{\sc iv}    & 13.7$\pm$0.8 & 12.25$\pm$0.02\\
\hline
\multicolumn{5}{c}{$z_{\rm abs} = 2.2298$} \\
\hline
2.22979  & 0.0 &  C~{\sc iv}   & 7.1$\pm$0.3   & 12.19$\pm$0.01\\
2.22978  & $-$1.0 &  O~{\sc vi}   & 19.3$\pm$0.9   & 13.61$\pm$0.02\\
2.23016  & +34.3 &  O~{\sc vi}   & 11.6$\pm$5.5   & 12.51$\pm$0.18\\
\hline

\end{tabular}
\end{table*}

\subsection{$z_{\rm abs}$ = 1.6574 }

Figure~A3 presents the velocity profiles and the VPFIT solutions for
the low and high ion species detected in this system. The H~{\sc i}
Ly$\alpha$ feature is strongly saturated over approximately
400~km~s$^{-1}$ (see Fig.~A3) with no damping wings. We therefore
did not attempt to derive highly uncertain H~{\sc i} column
densities. For the low ion species, we identified 9 distinct
components based on the fits to the Mg~{\sc ii} doublet, Si~{\sc
ii}${\lambda}$1526 and Al~{\sc ii}${\lambda1670}$ transitions. A
simultaneous fit was conducted on other low ion species by fixing
their redshifts and $b$ parameters to the values obtained above. The
results of this fit are presented in Table~A3. The two components
with the redshifts of $z = 1.6574$ and 1.6579 (components no. 5 and
8) are the strongest components in this system and it is only for
these two components that the O~{\sc i} and Mg~{\sc i} absorption
are detected. Of the 5 detected Si~{\sc ii} transitions, only
Si~{\sc ii}$\lambda1526$ is not suffering from saturation or
blending. Therefore, this is the only transition we use to derive
Si~{\sc ii} column densities. The Fe~{\sc ii}$\lambda2586$ and
$\lambda2600$ transitions are also contaminated by atmospheric
absorption lines and were not included in the Fe~{\sc ii} fit. We
could not identify Si~{\sc iii}$\lambda1206$ because it is
redshifted at the exact position of the DLA absorption of the system
at $z_{\rm abs} = 1.6359$. The Al~{\sc iii} doublet is also detected
but the red wing of the Al~{\sc iii}$\lambda1862$ transition is
blended with some unidentified absorption. It is interesting to note
that the Al~{\sc iii} component at velocity $v = +54$ km~s$^{-1}$ is
weak when this is the second strongest component in all the other
low ion species. Apart from this discrepancy, the Al~{\sc iii}
profile does track the low ion profiles closely enough to yield a
successful simultaneous solution.

The VPFIT solutions and velocity profiles of the C~{\sc iv}  and
Si~{\sc iv} doublets are also given in Fig.~A3. The two species were
fitted individually. Due to the very large velocity spread of this
system ($\approx 650$ km~s$^{-1}$), part of the C~{\sc
iv}$\lambda1548$ profile is blended with the C~{\sc iv}$\lambda1550$
profile. In Fig.A3, the blue shaded area in the C~{\sc
iv}$\lambda1548$ velocity panel (resp. C~{\sc iv}$\lambda1550$
velocity panel) indicates absorption from the C~{\sc
iv}$\lambda1550$ (resp. C~{\sc iv}$\lambda1548$). The Si~{\sc
iv}$\lambda1393$ absorption is heavily blended with the absorption
lines of the forest and we chose not to include it in the fit. The
fits to the C~{\sc iv} and Si~{\sc iv} doublets were successful with
17 and 9 individual components, respectively. It is worth noting
that due to the severe blending of the Si~{\sc iv} profile, our
attempt to tie the C~{\sc iv} and Si~{\sc iv} absorption profiles
failed, implying that the derived Si~{\sc iv} column densities might
not be so accurate. The velocity regions where the N~{\sc v} doublet
is expected are also depicted in Fig.~A3. We could not convince
ourselves that any significant N~{\sc v} absorption is present
although some faint feature can be seen around $+50$~km~s$^{-1}$.
Table~A3 lists the redshift, $b$ value and column density along with
1~$\sigma$ error of every velocity component from the VPFIT fit of
the high ion species.

\subsection{$z_{\rm abs}$ = 1.7221, 1.7536}

These two absorption systems are very weak and identified solely by
the presence of the C~{\sc iv} and  Ly$\alpha$ absorption (see
Fig.~A3). The expected Mg~{\sc ii} doublets of these absorbers are
either not detected or lost in the atmospheric absorption lines. The
parameters of the fits are listed in Table~A4.

\subsection{$z_{\rm abs}$ = 1.8994 }

This is another system identified only by the presence of Ly$\alpha$
and the C~{\sc iv} doublet. No other low or high ion species is
detected. The velocity profiles and VPFIT solutions for this C~{\sc
iv} system are presented in Fig.~A4 together with the wavelength
ranges where Si~{\sc iv} doublet absorption are expected. We found
12 components were required for an optimal fit to the C~{\sc iv}
absorption profiles. The results of the fit are presented in
Table~A4. This system extends over $\approx$400 km~s$^{-1}$ in
velocity space.

\subsection{$z_{\rm abs}$ = 1.9253 }

This absorption system is very weak and identified by the presence
of the C~{\sc iv} and  Ly$\alpha$ absorption (see Figs.~A4). The
expected Mg~{\sc ii} doublets of the absorber is either not detected
or lost in the atmospheric absorption lines. The parameters of the
fits are listed in Table~A4.

\subsection{$z_{\rm abs}$ = 2.2298 }

This system is recognized mainly by its C~{\sc iv} and O~{\sc vi}
doublets. The corresponding H~{\sc i} absorption profile is very
weak. However, a 3-component fit was conducted to the Ly$\alpha$
absorption profile (Fig.~A4) and yields log~$N_{\rm H I} = 12.75 \pm
0.08$ (orange curve). The absorption straddles two other H~{\sc i}
absorption profiles with log~$N_{\rm H I} = 13.38 \pm 0.01$ (blue
curve) and log~$N_{\rm H I} = 12.30 \pm 0.22$ (cyan curve).
Moreover, a single component fit to the C~{\sc iv} doublet was
optimal, yielding $\chi_{\nu}^2 =$ 1.09 and $P_{\chi^2} =$ 0.313.
The O~{\sc vi} doublet is suffering from blending and we performed a
2-component Voigt profile fit to the O~{\sc vi}$\lambda1037$
transition, which appears to be less blended than the O~{\sc
vi}$\lambda1031$. The fit was successful, yielding $\chi_{\nu}^2 =$
1.12 and $P_{\chi^2} =$ 0.308. Note that, due to the poor alignment,
the fit to different high-ions were performed separately. In
Fig.~A4, the VPFIT results are superimposed onto the observations as
red lines. Table~A4 gives the fit parameters.\\

Finally, Table~A5 presents the total column densities of the C~{\sc
iv} and Si~{\sc iv} species of the systems studied in this work. In
this table, columns 2 \& 3 indicate the velocity width of the C~{\sc
iv} and Si~{\sc iv} absorption features, respectively. The last two
columns also give the C~{\sc iv}/H~{\sc i} and Si~{\sc iv}/H~{\sc i}
column density ratios.

 \begin{table*}
  \centering
\caption{Total column densities of the C~{\sc iv} and Si~{\sc iv}
species of the systems studied in this work. In this table, column 1
shows the redshifts of the systems, columns 2 \& 3 indicate the
velocity width of the C~{\sc iv} and Si~{\sc iv} absorption
features, columns 3 \& 4 \& 5 indicate the total column densities of
C~{\sc iv}, Si~{\sc iv}, and H~{\sc i}, respectively, and finally,
the last two columns give the C~{\sc iv}/H~{\sc i} and Si~{\sc
iv}/H~{\sc i} column density ratios.}

\setlength{\tabcolsep}{6.8pt}
\renewcommand{\arraystretch}{1.0}
\begin{tabular}{ >{\scriptsize}c >{\scriptsize}c >{\scriptsize}c >{\scriptsize}c >{\scriptsize}c >{\scriptsize}c >{\scriptsize}c >{\scriptsize}c }

\hline \hline

  $z$    & $\Delta$V~(C~{\sc iv})~[km~s$^{-1}$] & $\Delta$V~(Si~{\sc iv})~[km~s$^{-1}$] & log~$N$(C~{\sc iv}) & log~$N$(Si~{\sc iv}) & log~$N$(H~{\sc i}) & log~$N$(C~{\sc iv})/$N$(H~{\sc i}) & log~$N$(Si~{\sc iv})/$N$(H~{\sc i}) \\

\hline

1.1846 &  79.6 &  79.6 & 13.98$\pm$0.04 & 13.68$\pm$0.06 &       ....       &      ....      & ....           \\
1.4681 &  21.9 &  .... & 12.89$\pm$0.01 &      ....      &       ....       &      ....      & ....            \\
1.5610 &  42.2 &  .... & 13.05$\pm$0.01 &      ....      &  14.13$\pm$0.01  & $-$1.08$\pm$0.01 & ....             \\
1.5965 & 234.4 & 175.5 & 15.08$\pm$0.02 & 13.28$\pm$0.02 &     $\ge$~17.85   &   $\le$ $-$2.77 &   $\le$ $-$4.57   \\
1.6109 & 160.8 & ....  & 14.06$\pm$0.01 &     ....       &     $\ge$~14.59   &   $\le$ $-$0.53 &   ....           \\
1.6359 & 187.6 & 187.6 & 13.54$\pm$0.09 & 12.81$\pm$0.05 &  19.15$\pm$0.04   & $-$5.61$\pm$0.09 & $-$6.34$\pm$0.06 \\
1.6574 & 551.5 & 157.9 & 14.84$\pm$0.02 & 14.18$\pm$0.01 &     ....          &     ....      &    ....        \\
1.6720 & 582.0 & 499.1 & 15.08$\pm$0.01 & 14.18$\pm$0.04 &  19.78$\pm$0.05   & $-$4.70$\pm$0.05 & $-$5.60$\pm$0.06 \\
1.7221 &  single-comp. &  .... & 12.43$\pm$0.01 &      ....      &  $\ge$~14.80  & $\le$ $-$2.37 & ....             \\
1.7536 &  26.1 &  .... & 12.89$\pm$0.02 &      ....      &  $\ge$~14.81  & $\le$ $-$1.92 & ....             \\
1.8994 & 355.5 &  .... & 13.70$\pm$0.02 &      ....      &       ....       &      ....      & ....            \\
1.9253 & single-comp.  &  .... & 12.25$\pm$0.02 &      ....      &  $\ge$~15.12  & $\le$ $-$2.87 & ....             \\
2.0422 &  131.0       &  .... & 13.70$\pm$0.02 &      ....      &  $\le$~12.90  & $\ge$ $+$0.80 & ....             \\
2.1979 & 28.1 & 7.5 & 13.24$\pm$0.02 & 12.23$\pm$0.04 &     $\ge$~15.98   &   $\le$ $-$2.74 &   $\le$ $-$3.72   \\
2.2028 & 43.1 & 43.1 & 14.14$\pm$0.06 & 13.33$\pm$0.06 &     $\ge$~15.54   &   $\le$ $-$1.40 &   $\le$ $-$2.21   \\
2.2298 & single-comp.  &  .... & 12.19$\pm$0.01 &      ....      &  12.75$\pm$0.08  & $-$0.56$\pm$0.08 & ....             \\
2.2363 & 48.2  &  .... & 12.66$\pm$0.04 &      ....      &  14.18$\pm$0.05  & $-$1.52$\pm$0.06 & ....             \\

\hline

\end{tabular}
\renewcommand{\footnoterule}{}
\end{table*}

\end{document}